\documentclass[journal]{IEEEtran}

\IEEEoverridecommandlockouts                             
\usepackage{graphicx} 
\usepackage{amsmath} 
\usepackage{amssymb}  
\usepackage{multicol}
\usepackage{epstopdf}
\usepackage{color}
\usepackage[usenames,dvipsnames,svgnames,table]{xcolor}
\usepackage{soul}
\usepackage[normalem]{ulem}
\usepackage{tikz}

\newtheorem{assumption}{Assumption}
\newtheorem{proposition}{Proposition}

\newtheorem{lemma}{Lemma}
\newtheorem{remark}{Remark}

\usepackage{color}

\def\calH{\mathcal{H}}
\def\bp{\mathbf{p}}

\newcommand\copyrighttext{%
  \footnotesize \textcopyright 2017 IEEE. Personal use of this material is permitted.
  Permission from IEEE must be obtained for all other uses, in any current or future 
  media, including reprinting/republishing this material for advertising or promotional 
  purposes, creating new collective works, for resale or redistribution to servers or 
  lists, or reuse of any copyrighted component of this work in other works.}
\newcommand\copyrightnotice{%
\begin{tikzpicture}[remember picture,overlay]
\node[anchor=south,yshift=10pt] at (current page.south) {\fbox{\parbox{\dimexpr\textwidth-\fboxsep-\fboxrule\relax}{\copyrighttext}}};
\end{tikzpicture}%
}

\newcommand{\col}{\mbox{col}}

\def\cale{{\cal E}}

\def\calw{{\cal W}}
\def\hal{{1 \over 2}}

\def\L2e{{\cal L}_{2e}}

\def\rea{\mathbb{R}}

\def\begequarr{\begin{eqnarray}}
\def\endequarr{\end{eqnarray}}
\def\begequarrs{\begin{eqnarray*}}
\def\endequarrs{\end{eqnarray*}}
\def\begarr{\begin{array}}
\def\endarr{\end{array}}
\def\begequ{\begin{equation}}
\def\endequ{\end{equation}}
\def\lab{\label}
\def\begdes{\begin{description}}
\def\enddes{\end{description}}
\def\begenu{\begin{enumerate}}
\def\begite{\begin{itemize}}
\def\endite{\end{itemize}}
\def\endenu{\end{enumerate}}

\def\lef[{\left[\begin{array}}
\def\rig]{\end{array}\right]}

\def\begcen{\begin{center}}
\def\endcen{\end{center}}
\def\begrem{\begin{remark}\rm}
\def\endrem{\end{remark}}

\def\calh{{\cal H}}
\def\calc{{\cal C}}

\def\cale{{\cal E}}

\def\rea{\mathbb{R}}

\def\begmat#1{\begin{bmatrix}#1\end{bmatrix}}
\def\begali#1{\begin{align}{#1}\end{align}}
\def\begalis#1{\begin{align*}{#1}\end{align*}}

\title{\LARGE \bf
New results on disturbance rejection for energy-shaping controlled port-Hamiltonian systems
}

\author{Joel Ferguson, Alejandro Donaire, Romeo Ortega and Richard H. Middleton 
\thanks{Joel Ferguson and Richard H. Middleton are with School of Electrical Engineering and Computing and PRC CDSC, The University of Newcastle, Callaghan, NSW, 2308, Australia.
        {\tt\small Email: Joel.Ferguson@uon.edu.au, Richard.Middleton@newcastle.edu.au}}
\thanks{Alejandro Donaire is with the Department of Electrical Engineering and Information Theory and PRISMA Lab, University of Naples Federico II, Napoli, 80125, Italy, and the School of Electrical Engineering and Computer Science, Queensland University of Technology, Brisbane, QLD, 4000, Australia. 
{\tt\small Email: Alejandro.Donaire@unina.it}}
\thanks{$^{3}$Romeo Ortega is with Laboratoire des Signaux et Syst\`emes, CNRS-SUPELEC, 91192, Gif-sur-Yvette, France
{\tt\small Email: ortega@lss.centralesupelec.fr}}
}

\begin{document}

\maketitle
\thispagestyle{empty}
\pagestyle{empty}
\copyrightnotice

\begin{abstract}
In this paper we present a method to robustify energy-shaping controllers for port-Hamiltonian (pH) systems  by adding an integral action that rejects unknown additive disturbances. The proposed controller preserves the pH structure and, by adding to the new energy function a suitable cross term between the plant and the controller coordinates, it avoids the unnatural coordinate transformation used in the past. This paper extends our previous work by relaxing the requirement that the systems Hamiltonian is strictly convex and separable, which allows the controller to be applied to a large class of mechanical systems, including underactuated systems with non-constant mass matrix. Furthermore, it is shown that the proposed integral action control is robust against unknown damping in the case of fully-actuated systems.
\end{abstract}
\section{INTRODUCTION}
\label{intro}

Port-Hamiltonian (pH) systems are a class of nonlinear dynamics that can be written in a form whereby the physical structure related to the system energy, interconnection and dissipation is readily evident \cite{Schaft2014}. Interconnection and damping assignment passivity-based control (IDA-PBC) is a control design that imposes a pH structure to the closed-loop dynamics \cite{Ortega2004}. This control method has been successfully applied to a range of nonlinear physical systems such as electrical machines \cite{Petrovic2001,BATetal}, power converters \cite{SEDetal}, chemical processes \cite{DORetal} and underactuated mechanical systems \cite{Acosta2005}, \cite{VIOetal}. Although IDA-PBC is robust to parameter uncertainty and passive preturbations, the presence of (practically unavoidable) external disturbances can significantly degrade its performance by shifting equilibria or even causing instability. In this paper, we focus on the problem of robustifying IDA-PBC {\em vis-\`a-vis} external disturbances via the addition of integral action control (IAC).

{A number of IACs have been proposed for pH systems, each conforming to specified control objectives. In \cite{Ortega2004}, an integrator is applied to the passive output of a pH system which has the effect of regulating the passive output to zero. This scheme can be interpreted as a control by interconnection (CbI), studied in  \cite{Ortega2007}, and is robust against parameters of the open-loop plant. A fundamentally different approach was taken in \cite{Donaire2009,Ortega2012} where the control objective was to regulate a signal that is not necessarily a passive output of the plant. As such, it was shown that the IAC for passive outputs proposed in \cite{Ortega2004} was not applicable as it cannot be implemented in a way that preserves the pH form. Rather, a partial state transformation was utilised to ensure that the closed-loop dynamics preserve the pH form. A critical step in this approach is the evaluation of the systems energy function at the transformed coordinates, which is a rather unnatural construction---see Remark 2 in  \cite{Donaire2016}.  This IAC was tailored for fully-actuated mechanical systems in \cite{Romero2013a} and underactuated mechanical systems in \cite{Donaire2016}. While in both cases the required change of coordinates to preserve the pH form were given explicitly, a number of technical assumptions were imposed to do so in the underactuated case. In both cases, the proposed IACs were shown to preserve the desired equilibrium of the system, rejecting the effect of an unknown matched disturbance.

More recently, an alternate approach to IAC design for pH systems was proposed in \cite{Ferguson2015,Ferguson}. The IAC is designed, as in \cite{Donaire2009,Ortega2012}, to regulate signals that are not necessarily passive outputs of the plant. The key advancement over previous solutions is that the IAC does not require any coordinate transformations in the design procedure. Rather, the energy function of the controller depends on both the states of the controller and the plant which allows preservation of the pH structure by construction. It has been shown in \cite{Ferguson} that the closed-loop can be interpreted as the power preserving interconnection of the plant and controller, which resembles the CbI technique \cite{Ortega2007}. It was also shown that, under a number of technical assumptions, the IAC could reject the effects of both a matched and unmatched disturbance.}

In this paper, the realm of application of the IAC presented in \cite{Ferguson} is significantly extended by relaxing some previously made assumptions. Specifically, the assumption in \cite{Ferguson} that the open-loop Hamiltonian is strongly convex and separable is not required here. By relaxing this assumption, our result can be applied to a broader class of pH system. In particular, our result applies to a strictly larger class of underactuated mechanical systems than the ones considered in \cite{Donaire2016} and \cite{Ferguson}.

The remainder of the paper is structured as follows: The problem formulation is presented in Section \ref{probform}. A brief summary of previous work is given in Section \ref{bg}. The new IAC scheme is presented in Section \ref{ia} and the stability properties of the closed-loop are considered in Section \ref{sec5}. The IAC is tailored for mechanical systems in Section \ref{Mech}. The control scheme is applied to three examples in Section \ref{ex}. Finally, the results of the paper are briefly discussed in Section \ref{con}.\\

\noindent {\bf Notation.}  Function arguments are declared upon definition and are omitted for subsequent use. For $x \in \rea^n$ we define $|x|^2=x^\top x$ and $\|x\|^2_P=x^\top P x$, for $P \in \rea^{n \times n}$, $P>0$. All functions are assumed to be sufficiently smooth.  For mappings $\mathcal{H}:\rea^n \to \rea$, $\mathcal{C}: \mathbb{R}^n \to \rea^{m}$ and $\mathcal{G}:\rea^p \times \rea^s \to \mathbb{R}$ we denote the transposed gradient as $\nabla\mathcal{H}:= \left(\frac{\partial \mathcal{H}}{\partial x}\right)^\top$, the transposed Jacobian matrix as $\nabla \calc:= \left(\frac{\partial \mathcal{C}}{\partial x}\right)^\top$ and $\nabla_{x}\mathcal{G}(x,y):= \left(\frac{\partial \mathcal{G}}{\partial x}\right)^\top$. For the distinguished $x^\star, \bar x \in \mathbb{R}^n$, we define the constant vectors $\bar \calc:=\calc(\bar x),\;\calc^\star:=\calc(x^\star)$, $\nabla \mathcal{H}^\star:= \nabla \mathcal{H}(x^\star)$ and $\nabla \bar{\mathcal{H}}:= \nabla \mathcal{H}(\bar{x})$. 

\section{Problem formulation}\label{probform}
\subsection{Perturbed system model}
\label{subsec21}

In this paper we consider the scenario where an unperturbed  system has been stabilised at a desired constant equilibrium using IDA-PBC and the objective is to add an IAC to reject additive disturbances. More precisely, the dynamics of the system are of the form:
\begequarr
\nonumber
		\begin{bmatrix}
		\dot x_a \\
		\dot x_u
		\end{bmatrix}
		&=&[J(x)-R(x)]
		\begin{bmatrix}
		\nabla_{x_a}\calH \\
		\nabla_{x_u} \calH
		\end{bmatrix} +
		\begin{bmatrix}
		u - d_a(x) \\ - d_u(x)
		\end{bmatrix} \\
		\nonumber
		y_a
		&=&
		\nabla_{x_a}\calH \\
		y_u
		&=&
		\nabla_{x_u}\calH,
\label{phdist}
\endequarr
where $x = \operatorname{col}(x_a, x_u)\in\mathbb{R}^n$ is the state vector, with $x_a \in \mathbb{R}^m$ and $x_u \in \mathbb{R}^s$, $s := n-m$, the actuated and unactuated states, respectively, $y_a \in \mathbb{R}^m$, $y_u \in \mathbb{R}^s$ are the signals to be regulated to zero and $u\in\mathbb{R}^m$ is the control input. The function $\mathcal{H}:\mathbb{R}^n \to \mathbb{R}_+$ is the Hamiltonian of the system. The interconnection and damping matrices are partitioned as   
\begin{equation}
	\begin{split}
		J(x) := \begin{bmatrix} J_{aa}(x) & J_{au}(x) \\ -J_{au}^\top(x) & J_{uu}(x) \end{bmatrix} &= -J^\top(x) \\
		R(x) := \begin{bmatrix} R_{aa}(x) & R_{au}(x) \\ R_{au}^\top(x) & R_{uu}(x) \end{bmatrix} &= R^\top(x)  \geq 0,
	\end{split}
\lab{jminr}
\end{equation}
respectively. The signals $d_a:\rea^n \to  \mathbb{R}^m$ and $d_u: \rea^n \to \mathbb{R}^s$ are the, state-dependent, matched and unmatched disturbances of the system, respectively.

We assume that the energy-shaping and damping injection steps of the IDA-PBC design have been accomplished for system \eqref{phdist}. This means, on one hand, that a desired equilibrium $x^\star \in \rea^n$ satisfies
$$
x^\star = \arg \min \calh(x)
$$
and is isolated, which implies that the system \eqref{phdist} without disturbances and $u=0_{m\times 1}$ has a stable equilibrium at $x^\star$. On the other hand, the damping injection step---consisting of a proportional feedback around the passive output---ensures that
$$
-\nabla^\top \calh \; R \; \nabla \calh \leq -\alpha |y_a|^2,
$$ 
and $\alpha >0$ is a constant. Consequently, the equilibrium is asymptotically stable if $y_a$ is a detectable output for the system. See \cite{Schaft2017} for further details on stability of pH systems.\\

Notice that we have taken the input matrix of the form $\begin{bmatrix} I_{m} \\ 0_{s\times m}\end{bmatrix}$. As is well-known \cite{NIJVAN} a necessary and sufficient condition to transform---via input and state changes of coordinates---an arbitrary input matrix into this form is that its columns span an involutive distribution.

\subsection{Assumptions }
\label{subsec22}
As indicated in the introduction the objectives of the IAC are to preserve the existence of a stable equilibrium and to ensure that the output signal ($y_a$ and $y_u$ when $d_u=0_{s\times 1}$ or $y_u$ if $d_a=0_{m\times 1}$) is driven to zero in spite of the presence of disturbances. Also, some stability properties should be preserved when both disturbances act simultaneously. In this subsection we present the assumptions that are imposed on the system \eqref{phdist} to attain these objectives. 

For the case of matched disturbances it is possible to preserve in closed-loop the original equilibrium $x^\star$. This, clearly, implies that if this equilibrium is asymptotically stable then $|y_a(t)| \to 0$, $|y_u(t)| \to 0$ as desired. In order to ensure the former property for state-dependent disturbances the following assumption is imposed.

\begin{assumption}\label{Assumption:MatchedDist}
	The disturbance $d_a$ can be written in the form
	$$
		d_a = G_d(x)\bar{d}_a,
	$$
	where $\bar d_a \in \mathbb{R}^m$ is constant and $G_d: \rea^n \to \mathbb{R}^{m\times m}$. Moreover, $G_d$ is full rank and sign definite. Without loss of generality, it is assumed that $G_d<0$.
\end{assumption}
 
Similarly to the case above, to handle the case of state-dependent unmatched disturbances it is necessary to assume they satisfy a structural condition, which is articulated as follows.

\begin{assumption}\label{Assumption:J12R12Const}
The disturbance $d_u$ can be written in the form
\begin{equation}
	d_u = (J_{au}+R_{au})^\top \bar d_u,
\end{equation}
where $\bar d_u \in \mathbb{R}^m$ is constant.
\end{assumption}

{An additional difficulty for the unmatched disturbance case is that  it is not possible to preserve the original equilibrium, even in the case when $d_u$ is constant---see \cite{Ortega2012} for a detailed discussion. Therefore, it is necessary to consider another value for $x$ in the closed-loop to be stabilized, that we denote $\bar x = (\bar x_a,\bar x_u) \in \rea^n$. The new equilibrium $\bar x$ should belong to the set
\begin{equation} \label{eqset}
	\begin{split}
		\cale:= \{x \in \rea^n| -(J_{au}+ R_{au})^\top (\nabla_{x_a}\calH + \bar d_u)\\+(J_{uu}-R_{uu}) \nabla_{x_u}\calH =0_{s\times 1} \},
	\end{split}
\end{equation}
which is the set of assignable equilibria. The following assumption guarantees the existence of such an $\bar x$ and is utilised later for stability analysis.}

\begin{assumption}\label{properAssump2}
There exists an isolated $\bar x \in \mathbb{R}^n$ satisfying
	\begin{equation}
		\bar x =\arg \min \mathbf{H}(x)
	\end{equation}
	where
	\begin{equation}
	\lab{bfh}
		\mathbf{H}(x) := \mathcal{H}(x) + x_a^\top \bar d_u.
	\end{equation}
\end{assumption} 

The following remarks regarding the assumptions are in order.
\begite
\item[{\bf R1.}] {Consistent with the internal model principle, the assertion that $\bar d_u \in \mathbb{R}^m$ in Assumption \ref{Assumption:J12R12Const} implies that there are enough control actions to reject the unmatched disturbance. See \cite{Ortega2012} for further discussions on the need for this condition.}

\item[{\bf R2.}] Unmatched disturbances of the form considered in Assumption \ref{Assumption:J12R12Const} can be equivalently described by a matched disturbance and a ``shifted" Hamiltonian. This property is utilised to analyse the effects of unmatched disturbances.

\item[{\bf R3.}] To verify that $\bar x$, is indeed an assignable equilibrium, notice that as it minimises the function $\mathbf{H}$, it satisfies $\nabla_{x_a}\bar{\mathcal{H}}=-\bar d_u, \nabla_{x_u}\bar{\mathcal{H}}=0_{m\times 1}$ which implies that $\bar x \in \cale$. Moreover, as $ \nabla_{x_u}\bar \calH =0_{s\times 1}$ , if the point $\bar x$ is asymptotically stable, then $|y_u(t)| \to 0$, which is part of the control objective.

\item[{\bf R4.}] The particular form of $\mathbf{H}$ in Assumption \ref{properAssump2} is necessary to construct a Lyapunov function to study the ``shifted" equilibrium $\bar x$. Clearly, the assumption is satisfied if $\mathcal{H}$ is convex (at least locally in the domain of interest).
\endite
\subsection{Problem statement}
\label{subsec23}
Consider the pH system \eqref{phdist} verifying Assumption \ref{Assumption:MatchedDist} when $d_u=0_{s\times 1}$ and Assumptions \ref{Assumption:J12R12Const}-\ref{properAssump2} when $d_a=0_{m\times 1}$. Define mappings $\hat u: \rea^n \times \rea^m \to \rea^m$ and $F:\rea^n \to \rea^m$ such that the  IAC 
\begalis{
u&=\hat u(x,{x}_c)\\
\dot {x}_c &=F(x)
}
ensures the closed-loop is an, unperturbed, pH system with an (asymptotically) stable equilibrium at $(x^\star,\bar {x}_c)$ when $d_u=0_{s\times 1}$ and at $(\bar x,\bar {x}_c)$ when $d_a=0_{m\times 1}$, for some $\bar {x}_c \in \rea^m$. Moreover, give conditions under which a stable equilibrium exists in the presence of, both, matched and unmatched disturbances.\\

%
\section{Previous Work and Contributions of the Paper}\label{bg}
%
\subsection{Integral action on passive outputs $y_a$}\label{subsec31}

For the case of constant, matched disturbances $\bar d_a$ it is well-known \cite{Ortega2004,Ortega2012} that adding an IAC around the passive outputs $y_a$ of the form
\begin{equation}\label{IntActionCtrl}
	\begin{split}
		\dot{x}_c &= K_i \; y_a \\
		u &= -{x}_c,
	\end{split}
\end{equation}
where $K_i>0$, ensures the closed-loop is a pH system with a stable equilibrium at $(x^\star,\bar d_a)$ and guarantees that $|y_a(t)| \to 0$. The equilibrium is, moreover, asymptotically stable if $y_a$ is a detectable output, which ensures that the non-passive output $y_u$, also converges to zero. 

Unfortunately, the detectability condition is rather restrictive, hence, the need to propose alternative IACs even in the case $d_u=0_{s\times 1}$. In fact, the IAC \eqref{IntActionCtrl} cannot ensure detectability when used for mechanical systems \cite{Romero2013a}. Moreover, this simple output-feedback construction is applicable only to the passive output. Indeed, it is shown in \cite{carlesrobust} that it is not possible to use an IAC of the form \eqref{IntActionCtrl} around $y_u$ preserving the pH form---see also the discussion in  \cite{Donaire2009}.

\subsection{Integral action of $y_u$ via coordinate transformations}
\lab{subsec32}
To reject unmatched disturbances it seems reasonable to add an IAC around the output $y_u$. An approach to carry out this task was proposed in \cite{Donaire2009} and further investigated in \cite{Ortega2012,Romero2013a,Donaire2016}. The key step in those papers is the solution of a nonlinear algebraic equation that ensures the existence of a change of coordinates
\begin{equation}\label{zTranform}
	z=\col(z_1,z_2,z_3)=\col(\psi(x_a,x_u,{x}_c),x_u,{x}_c)
\end{equation}
and an IAC
\begali{
\nonumber
u&=\hat u(x,{x}_c)\\
\lab{iacort}
\dot {x}_c &=K_i \nabla_{x_u} \calh(\psi(x_a,x_u,{x}_c),x_u)
}
such that, the closed-loop system written in the new coordinates has a pH structure. 
\begequ
\lab{phsysz}
\dot z = \lef[{cc} \left.[J(x)-R(x)]\right|_{x=(\mu(z),z_2)} & \lef[{c}  0_{m\times s} \\ -K_i^\top \rig] \\ \lef[{cc} 0_{s\times m} & K_i \rig] & 0_{s\times s} \rig]\nabla \mathcal{H}_{cl}(z)
\endequ
where $J$ and $R$ are the open-loop interconnection and damping matrices defined in \eqref{jminr}, $\mu:\rea^{n+m} \to \rea^m$ is a left inverse of $\psi$ in the sense that 
$$
\psi(\mu(z),z_2,z_3)=z_1,
$$ 
and $\calh_{cl}$ is the new Hamiltonian function given by
$$
\mathcal{H}_{cl}(z) = \mathcal{H}(z_1,z_2) + \frac12 \|z_3-\bar d_u\|^2_{K_i^{-1}}.
$$ 
As discussed in Subsection \ref{subsec22} in the case of unmatched disturbances it is not possible to preserve the open-loop equilibrium of the unperturbed system $x^\star$. To ensure that an equilibrium exists for the IAC \eqref{iacort} it is necessary to impose an assumption, similar to Assumption \ref{properAssump2} (see remark R3. of Subsection \ref{subsec22}), namely  
\begequ
\lab{barxiacort}
\exists\; \hat x \in \cale \cap\{ x \in \rea^n\;|\;\nabla_{x_u} \calh(\psi(x_a,x_u,\bar d_u),x_u)=0_{s \times 1}\},
\endequ
with $\cale$ defined in \eqref{eqset}.
Under this assumption it can be shown that the closed-loop system \eqref{phsysz} has a stable equilibrium at 
$$
\bar z:=(\psi(\hat x_a, \hat x_u,\bar d_u),\hat x_u,\bar d_u).
$$
Furthermore, if the output $y_a$ is detectable the equilibrium is asymptotically stable and $|y_u(t)| \to 0$. {Interestingly, it can be shown that if the system \eqref{phdist} satisfies Assumption \ref{properAssump2} and there exists a suitable transformation \eqref{zTranform}, then the condition \eqref{barxiacort} is satisfied---the proof of this fact is provided in Lemma \ref{AssumpRelation} in the Appendix.}

{In addition to the above IAC, the coordinate transformation \eqref{zTranform} has been utilised for damping injection into unactuated coordinates (see \cite{Romero2013a} and \cite{Donaire2016}). This extension has allowed the construction of strict Lyapunov functions and verification of additional stability properties such as exponential stability and input-to-state stability.}

In spite of the interesting stability properties of the IAC \eqref{iacort} quoted above, there are several limitations of this approach. First, the solution of the aforementioned algebraic equation---which should satisfy some restrictive rank and matching conditions---is often problematic. Interestingly, it has been solved for permanent magnet synchronous motors (PMSM) in \cite{Donaire2009} and for fully actuated and a class of underactuated mechanical systems in \cite{Romero2013a} and \cite{Donaire2016}, respectively. Second, as indicated in the introduction, the construction of the new Hamiltonian function is rather unnatural and, in contrast with the technique proposed in this paper, is not amenable for a physical interpretation of the IAC. Finally, as remarked in \cite{Romero2013a} and \cite{Donaire2016}, the resulting expressions for the controllers are, in general, quite involved.
  
\subsection{Integral action of $y_u$ without coordinate transformations}
\lab{subsec33}
An alternate approach to apply IAC to $y_u$ without using coordinate transformations was proposed in \cite{Ferguson}. The paper considers a plant of the form \eqref{phdist} with $R_{au} = 0_{m\times s}$, constant disturbances and $\mathcal{H}$ strongly convex and separable. The control objective is to design an IAC that preserves in closed-loop the component of the original equilibrium associated to the unactuated coordinate, that is, the desired closed-loop equilibrium is of the form $(\bar x_a, x_u^\star,\bar {x}_c)$. Under the assumption of separability of $\calh$ the asymptotic stability of this equilibrium implies $|y_u(t)| \to 0$.   

The IAC proposed in \cite{Ferguson} is given by 
\begin{equation}
	\begin{split}
		u &= (J_{aa} - R_{aa})\nabla_{x_a}\mathcal{H}_c \\
		\dot{x}_c &= E^\top J_{au}\nabla_{x_u}\mathcal{H},
	\end{split}
\end{equation}
where ${x}_c \in \mathbb{R}^s$, $E\in \mathbb{R}^{m\times s}$ is a constant matrix and $\mathcal{H}_c:\rea^s \to \rea$ is a strictly convex function with argument $E^\top x_a - {x}_c$. Thanks to this particular choice of argument of $\calh_c$ it is possible to prove that the closed-loop dynamics is a pH system of the form
\begin{equation}\label{OldCLdynamics}
	\begin{split}
		\begin{bmatrix}
		\dot{x}_a \\
		\dot{x}_u \\
		\dot{x}_c
		\end{bmatrix}
		&= 
		\begin{bmatrix}
		J_{aa} - R_{aa} & J_{au} & 0_{m\times s} \\
		-J_{au}^\top & J_{uu} - R_{uu} & -J_{au}^\top E\\
		0_{s\times m} & E^\top J_{au} & 0_{s\times s}
		\end{bmatrix} \nabla \calh_{cl} \\ 
		&\phantom{---}-
		\begin{bmatrix}
		d_a & d_u & 0_{s\times 1}
		\end{bmatrix}^\top.
	\end{split}
\end{equation}
with 
$$
\calh_{cl}(x,{x}_c):=\calh(x_a,x_u)+\calh_c(E^\top x_a - {x}_c),
$$
which has the property that
$$
\nabla \calh_{cl}=\begin{bmatrix}
		\nabla_{x_a}\mathcal{H} + E \nabla \mathcal{H}_c \\
		\nabla_{x_u}\mathcal{H} \\
		-\nabla\mathcal{H}_c
		\end{bmatrix}.
$$
{It was shown that if $J_{aa} - R_{aa}$ is constant, the desired closed-loop equilibrium is asymptotically stable for any constant matched disturbance $\bar d_a$. In the case of unmatched disturbances, it was shown that the desired closed-loop equilibrium is asymptotically stable provided that there exists a suitable equilibrium for the system, $J_{au}$ is constant and $\bar d_u$ is in the range of $J_{au}^\top$. When matched and unmatched disturbances coexist, stability of the closed-loop was guaranteed and $|y_u(t)| \to 0$, provided that a suitable equilibrium exists and the matrices $J_{aa} - R_{aa}$ and $J_{au}$ are constant.}
\subsection{Integral action control for mechanical systems}
\lab{subsec34}
Applying an IDA-PBC to a perturbed mechanical systems yields a closed-loop of the form \cite{Acosta2005}
\begin{equation}\label{mecdist}
	\begin{split}
		\begin{bmatrix}
	         \dot{q} \\
	         \dot{\bp}
		\end{bmatrix}
		&=
		\begin{bmatrix}
	          0_{{\ell}\times {\ell}} & M^{-1}(q)\mathbf{M}_d(q) \\
	         -\mathbf{M}_d(q)M^{-1}(q) & \mathbf{J}_2(q,\bp)-R_d(q)
		\end{bmatrix}
		\nabla \mathbf{H}_d \\
		&\phantom{---}+
		\begin{bmatrix}
		    d_u \\
		    G(q)(u-d_a)
		\end{bmatrix}
	\end{split}
\end{equation}
where $q,\mathbf{p} \in \mathbb{R}^{\ell},\; \ell:={n \over 2}$ are the generalised coordinate and momentum vectors,  $G$ is the full rank input matrix,  $M$ and $\mathbf{M}_d$ are the positive definite inertia matrices of the open and closed-loop, respectively, $R_d \geq 0$ is the dissipation matrix, $\mathbf{J}_2=- \mathbf{J}^\top_2$ is a designer chosen matrix capturing the effect of gyroscopic forces, $\mathbf{H}_d$ is the desired Hamiltonian function, which is of the form
\begin{equation}
\mathbf{H}_d(q,\mathbf{p})= \frac 12 \|\bp\|^2_{\mathbf{M}_d^{-1}(q)} + V_d(q),
\end{equation}
where $V_d$ is the closed-loop potential energy.

IDA-PBC ensures that $V_d$ has an isolated minimum at a desired value $q^\star$. The latter fact, together with positivity of the matrix $ \mathbf{M}_d$, implies that $(q^\star,0)$ is a stable equilibrium of \eqref{mecdist} when $u,d_a$ and $d_u$ are zero. The robust regulation control objective is to design an IAC that ensures stability of the desired equilibrium  $(q^{\star},0,\bar {x}_c)$.

In \cite{Romero2013a} it is proven that the addition of an IAC of the form \eqref{IntActionCtrl} to mechanical systems has catastrophic effects---making the achievement of the control objectives a zero measure event. IACs using the coordinate transformation technique of Subsection \ref{subsec32} have been reported for fully actuated systems in  \cite{Romero2013a} and extended in \cite{Donaire2016} to underactuated systems with $M_d$ constant and  
\begequ
\lab{assijrnlc}
G^\perp(q)\nabla_q(p^\top M^{-1}(q)p)=0,
\endequ
where $G^\perp:\rea^{\ell} \to \rea^{(\ell - m) \times m}$ is a full-rank left annihilator of $G$. In both papers, stability of the desired equilibrium is ensured in the absence of unmatched disturbances. Notice that Assumption \eqref{assijrnlc} implies that the inertia matrix is independent of the non-actuated coordinates that, together with the condition of constant $M_d$, rules out many systems of practical interest. More recently, the approach of Subsection \ref{subsec33} has been used in \cite{Ferguson} assuming full actuation and constant inertia matrices.

%
\subsection{Contributions of the paper}
\lab{subsec35}
The main contributions of the present paper may be summarized as follows:
\begite
\item[{\bf C1.}] Compared with the IAC of Subsection \ref{subsec31}: (i) the method is applicable for both matched and unmatched disturbances; and {(ii) the restrictive assumption of detectability of $y_a$ is significantly relaxed by the addition of damping between the actuated coordinates and the dynamic extension.}  
\item[{\bf C2.}] Compared with the IAC of Subsection \ref{subsec32}: (i) the need to solve a nonlinear algebraic equation is obviated; (ii) the construction of the closed-loop Hamiltonian function is more natural, and admits a physical interpretation; and (iii) the resulting IACs are considerably simpler.
\item[{\bf C3.}] Compared with the IAC of Subsection \ref{subsec33}: (i) the assumption of convexity and separability of the systems Hamiltonian $\calh$ are removed; {(ii) the result is extended to a class of state-dependent matched and unmatched disturbances;} and (iii) the detectability assumption required for asymptotic stability is relaxed---see point (ii) of C1 above.
\item[{\bf C4.}] Compared with the IACs for mechanical systems of Subsection \ref{subsec34} the assumptions of  $M_d$ constant and \eqref{assijrnlc} are removed, significantly enlarging the class of underactuated mechanical systems for which the method is applicable. Moreover, the resulting  IACs are much simpler.
\endite

%
\section{New Closed-Loop Port-Hamiltonian Structure}\label{ia}
In this section we present a novel IAC that does not rely on changes of coordinates nor assumes separability of the systems Hamiltonian. This IAC generates a new pH closed-loop system that, with a suitable selection of the tuning gains, provides the solutions to the problems formulated in Subsection \ref{subsec23}. 

\subsection{New closed-loop pH dynamics} \label{subsec41}

\begin{proposition}\label{propiacl} 
Consider the system \eqref{phdist} in closed-loop with the controller 
\begin{eqnarray} 
		u &=& [-J_{aa}+R_{aa}+J_{c_1}(x)-R_{c_1}(x)-R_{c_2}(x)]\nabla_{x_a} \calH \nonumber \\ 
		&&\phantom{--} + [J_{c_1}(x)-R_{c_1}(x)]{K_i}( x_a- {x}_c) + 2 R_{au} \nabla_{x_u} \calH \nonumber \\
		\dot{x}_c&=&-R_{c_2}(x)\nabla_{x_a}\mathcal{H} + (J_{au} + R_{au}) \nabla_{x_u}\mathcal{H}, \label{controlLaw2}
\end{eqnarray}
where ${x}_c \in \mathbb{R}^m$ and the (possibly state-dependent) matrices $J_{c_1}=-J_{c_1}^\top$, $R_{c_1} > 0$ and $R_{c_2} \geq 0$ are chosen by the designer. Then, the closed-loop dynamics expressed in the coordinates
$$
w = \begmat{w_a \\ w_u \\ w_c } := \begmat{x_a\\ x_u \\ x_a-{x}_c},
$$
can be written in the pH form
\begin{equation}\label{iacl}
		\dot w
		=
		\left[ J_{cl}(w) - R_{cl}(w) \right]
		\nabla\mathcal{H}_{cl}(w)
		-
		\begin{bmatrix}
			d_a \\ d_u \\ d_a
		\end{bmatrix},
\end{equation}
with new interconnection and damping matrices given by
\begequarr
\nonumber
J_{cl}&:=&  
\begin{bmatrix}
J_{c_1} & J_{au} + R_{au} & J_{c_1} \\
- (J_{au}+ R_{au})^\top & J_{uu} &  0_{s\times m}\\
 J_{c_1} & 0_{s\times s} & J_{c_1}
\end{bmatrix}\\
\label{Rcl}
R_{cl}&:= & 
\begin{bmatrix}
R_{c_1} + R_{c_2} & 0_{m\times s} & R_{c_1} \\
0_{s\times m} & R_{uu} & 0_{s\times m} \\
R_{c_1} & 0_{m\times s} & R_{c_1} 
\end{bmatrix},
\endequarr
and $\mathcal{H}_{cl}:\mathbb{R}^{2m+n}\to\mathbb{R}$ is the closed-loop Hamiltonian defined as 
\begin{equation}
\mathcal{H}_{cl}(w):= \mathcal{H}(w_a,w_u) + \hal\|w_c\|^2_{K_i}.
\end{equation}
\end{proposition}
%
\begin{IEEEproof} 
First, we underscore the presence of two key terms in the control signal \eqref{controlLaw2}: first, the term $-(J_{aa}-R_{aa})\nabla_{x_a} \calH$ than cancels the $(1,1)$-block of the open-loop system matrix; second, the term $2 R_{au} \nabla_{x_u} \calH$ that changes the sign of the of the $(1,2)$-block of the open-loop dissipation matrix. Hence, the dynamics of $x_a$ reduces to
\begali{
\nonumber
\dot x_a  =& (J_{c_1}-R_{c_1}-R_{c_2})\nabla_{x_a} \calH +(J_{au} + R_{au}) \nabla_{x_u} \calH\\
\lab{dotxa}
 & + (J_{c_1}-R_{c_1})K_i(x_a -{x}_c) - d_a.
}
Now, notice that
\begin{equation}\label{ianablarel}
	\nabla \calh_{cl}=\begmat{\nabla_{x_a}\mathcal{H} \\\nabla_{x_u}\mathcal{H} \\ K_i (x_a - {x}_c)},
\end{equation}
which replaced in \eqref{dotxa} yields
\begalis{
\dot x_a  =& (J_{c_1}-R_{c_1}-R_{c_2})\nabla_{w_a} \calH_{cl} +(J_{au} + R_{au}) \nabla_{w_u} \calH_{cl}\\
 & + (J_{c_1}-R_{c_1})\nabla_{w_c} \calH_{cl} - d_a,
}
which is the dynamics of $w_a$ in  \eqref{iacl}. 

The proof follows noticing that the second rows of \eqref{phdist} and \eqref{iacl} match. Finally, writing the controller dynamics \eqref{controlLaw2} as
$$
		\dot{x}_c =-R_{c_2} \nabla_{w_a}\calH_{cl} + (J_{au} + R_{au}) \nabla_{w_u}\calH_{cl}
$$
and substracting it from the expression of $\dot x_a$ above yields the last row of \eqref{iacl}. Thus, the closed-loop dynamics can be written in the pH form \eqref{iacl} as claimed.
\end{IEEEproof}

\subsection{Discussion} \label{subsec42}
The following remarks are in order.

\begite
\item[{\bf R5.}] The significance of the new IAC is easily appreciated comparing the closed-loop damping matrices of the new IAC \eqref{Rcl} with the ones of the IACs of Subsections \ref{subsec32} and \ref{subsec33}, namely, \eqref{phsysz} and  \eqref{OldCLdynamics}, respectively. While the two latter IACs leave this matrix unaltered (with respect to the open-loop system) the new IAC adds, via $R_{c_1}$ and $R_{c_2}$, damping to the new dissipation matrix. Notice that this is achieved ``swapping" the $(1,2)$ block $R_{au}$ to the interconnection matrix  \eqref{Rcl}. This additional damping injection is, of course, fundamental to enhance the performance of the IAC.
 
\item[{\bf R6.}] 
	For simplicity we have proposed the quadratic function  $\hal\|w_c\|^2_{K_i}$ as the energy of the controller dynamics. Proposition \ref{propiacl}, and subsequent results of the paper, can be easily extended to any convex function  $H_c(w_c)$, replacing in \eqref{controlLaw2}  the term ${K_i}( x_a- {x}_c)$ by $\nabla H_c$.  

\item[{\bf R7.}] 
As indicated in the proof, the control signal in \eqref{controlLaw2} contains a partial linearizing term $(-J_{aa}+R_{aa})\nabla_{x_a} \calH$ and replaces the $(1,1)$ block of the $J-R$ matrix by $J_{c_1}-R_{c_1}-R_{c_2}$. The first two terms are introduced to be able to satisfy Assumption \ref{Assumption:MatchedDist}, while the third one adds damping when $d_u=0$.

\item[{\bf R8.}] 
A drawback of the proposed controller is that it requires the knowledge of open-loop dissipation matrix $R$ that, in general, is uncertain. In the case that $R_{aa}$ is constant and positive definite and $R_{au} = 0_{m\times s}$, the IAC \eqref{controlLaw2} can be made independent of $R$. Indeed, by choosing the controller gains as
\begin{equation}
	\begin{split}
		J_{c_1} &= 0_{m\times m} \\
		R_{c_1} &= R_{aa} \\
		K_i &= \kappa R_{aa}^{-1},
	\end{split}
\end{equation}
where $\kappa > 0$ is a tuning parameter, the IAC simplifies to
\begin{eqnarray} 
		u &=& [-J_{aa}-R_{c_2}(x)]\nabla_{x_a} \calH -\kappa(x_a-{x}_c) \nonumber \\
		\dot{x}_c&=&-R_{c_2}(x)\nabla_{x_a}\mathcal{H} + J_{au}\nabla_{x_u}\mathcal{H},
\end{eqnarray}
which is clearly independent of the open-loop damping $R$.
{
\item[{\bf R9.}] 
The IAC \eqref{controlLaw2} can be equivalently expressed with integrator state $w_c$ as
\begin{equation}\label{controlLaw3}
	\begin{split}
		u &= [-J_{aa}+R_{aa}+J_{c_1}-R_{c_1}-R_{c_2}]\nabla_{x_a} \calH \\ 
		&\phantom{--} + [J_{c_1}-R_{c_1}]K_iw_c + 2 R_{au} \nabla_{x_u} \calH \\
		\dot{w}_c&=\left[J_{c_1}-R_{c_1}\right]\left[\nabla_{x_a}\mathcal{H} + K_iw_c\right]-d_a, 
	\end{split}
\end{equation}
In the case of unmatched disturbances only ($d_a=0_{m\times 1}$) the IAC can be implemented in this form. This may be advantageous in applications where $R_{au}=0_{m\times s}$ as the IAC no longer has dependency on $y_u=\nabla_{x_u}\mathcal{H}$, the signal to be regulated.}
\endite

%
\section{Main Results}\label{sec5}

In this section, the stability properties of the closed-loop dynamics \eqref{iacl} subject to matched and unmatched disturbances are studied. To simplify the presentation we consider first, in the next two subsections, each of these cases separately and then, in Subsection \ref{subsec44}, we treat the case where both disturbances are present.

The analysis in all cases proceeds along the following lines. First, we verify that the conditions imposed on the disturbances by Assumptions \ref{Assumption:MatchedDist} and \ref{Assumption:J12R12Const} ensure, after a suitable choice of the free parameters ($J_{c_1},R_{c_1}$ and $R_{c_2}$) of the IAC \eqref{controlLaw2}, the existence of an equilibrium $\bar w \in \rea^{n+m}$ for the closed-loop \eqref{iacl}. Second, establish the stability of the equilibrium using the shifted Hamiltonian function proposed in \cite{Jayawardhana2007}, that is,
\begequ
\lab{shiham}
\calh_{cl}(w)-\nabla^\top \bar \calh_{cl}(w - \bar w)-\bar \calh_{cl},
\endequ  
as a candidate Lyapunov function. Finally, looking at the time derivative of this function identify the detectability conditions that ensure asymptotic stability.
%
\subsection{The case of matched disturbances} \label{subsec51}
Consider the closed-loop dynamics \eqref{iacl} with $d_u = 0_{s\times 1}$ and the control gains 
\begin{eqnarray}\label{Jc1Rc1}
J_{c_1} = \frac 12 (G_d-G_d^\top), \; R_{c_1} = -\frac 12 (G_d+G_d^\top), \; R_{c_2} > 0. \label{gains}
\end{eqnarray}
This selection implies that 
\begequ
\lab{gd}
J_{c_1} - R_{c_1} = G_d.
\endequ

\begin{proposition}\label{propmatched}
	Consider the system \eqref{phdist} with $d_u=0_{s\times 1}$ and $d_a$ satisfying Assumption \ref{Assumption:MatchedDist}. Let the IAC be given by \eqref{controlLaw2} with the controller parameters selected as in \eqref{gains}. 
\begin{itemize}
\item[(i)] The $(n+m)$-dimensional vector
\begin{equation}\label{matchedEquilibrium}
	\bar w := (x_a^\star,x_u^\star,K_i^{-1}\bar d_a)
\end{equation}
is a stable equilibrium of the closed-loop system. 
\item[(ii)] If the signal
\begin{equation}\label{detectOutputMatched}
	Y_a:=\begin{bmatrix} \nabla_{w_a}\mathcal{H}_{cl} \\ \tilde w_c \end{bmatrix},
\end{equation}
where $\tilde {(\cdot)}:=(\cdot)-\bar{(\cdot)}$, is a detectable output, the equilibrium is asymptotically stable.
\item[(iii)] The stability properties are global if the Hamiltonian function $\calh$ is radially unbounded. 
\end{itemize}
\end{proposition}

\begin{IEEEproof}
To verify that \eqref{matchedEquilibrium} is an equilibrium of  \eqref{iacl} with $d_u = 0_{s\times 1}$ note that from \eqref{ianablarel}  we get
$$
\nabla \bar \calh_{cl} =\begmat{ 0_{m\times 1} \\ 0_{s\times 1}\\ \bar d_a},
$$ 
since  $x^\star$ is a minimum of $\mathcal{H}$. Replacing this equation in  \eqref{iacl} with $d_u = 0_{s\times 1}$ we get
\begalis{
\dot w|_{w = \bar w} & = \begmat{(\bar J_{c_1} - \bar R_{c_1})K_i \bar w_c- d_a(\bar x) \\ 0_{s\times 1}\\ (\bar J_{c_1} - \bar R_{c_1})K_i \bar w_c - d_a(\bar x) }\\
 & = \begmat{\bar G_d(K_i \bar w_c - \bar d_a) \\ 0_{s\times 1}\\ \bar G_d(K_i \bar w_c - \bar d_a)}=0_{(n+m)\times 1}
}
where we have invoked Assumption \ref{Assumption:MatchedDist} and \eqref{gd} to get the second identity and $K_i \bar w_c = \bar d_a$ for the third one. 

To prove that \eqref{matchedEquilibrium} is stable we evaluate the shifted Hamiltonian function \eqref{shiham} and define
$$
\mathcal{W}(w) :=\mathcal{H}_{cl}(w_a,w_u) - \bar d_a^\top(w_c-\bar w_c)-\bar \calh_{cl},
$$
which is clearly positive-definite, therefore qualifies as a Lyapunov candidate for the closed-loop system.  The time derivative of $\mathcal{H}_{cl}$ verifies
\begalis{
		\dot{\mathcal{H}}_{cl}
		=&
		\nabla^\top\mathcal{H}_{cl}(J_{cl} - R_{cl})\nabla\mathcal{H}_{cl} - (\nabla_{w_a}\mathcal{H}_{cl}+\nabla_{w_c} \calh_{cl})^\top d_a \\
		=&
		-\|\nabla_{w_a}\mathcal{H}_{cl}\|_{R_{c_2}}^2 +\|\nabla_{w_a}\mathcal{H}_{cl}+K_i w_c\|_{G_{d}}^2 \\
		&-\|\nabla_{w_u}\mathcal{H}_{cl}\|_{R_{uu}}^2-(\nabla_{w_a}\mathcal{H}_{cl}+K_i w_c)^\top G_{d} \bar d_a\\
		\leq&
		-\|\nabla_{w_a}\mathcal{H}_{cl}\|_{R_{c_2}}^2 +\|\nabla_{w_a}\mathcal{H}_{cl}+K_i w_c\|_{G_{d}}^2 \\
		&-(\nabla_{w_a}\mathcal{H}_{cl}+K_i w_c)^\top G_{d} \bar d_a.
}
where we used the fact that $R_{uu}\geq0$ to get the inequality. Likewise, the time derivative of the term $\bar d_a^\top w_c$ satisfies
\begalis{
		\bar d_a^\top \dot w_c
		&=
		\bar d_a^\top G_d(\nabla_{w_a}\mathcal{H}_{cl}+K_i w_c-\bar d_a).
}
Combining these two equations we complete the squares and get the bound 
\begin{equation}\label{LaypTimeDer}
		\dot{\mathcal{W}} \leq -\|\nabla_{w_a}\mathcal{H}_{cl}\|_{R_{c_2}}^2 -\|\nabla_{w_a}\mathcal{H}_{cl}+K_i \tilde w_c \|_{R_{c_1}}^2 \leq 0,
\end{equation}
which proves claim (i).

The proof of claim (ii) is established with LaSalle's invariance principle and the following implication
$$
\dot \calw = 0\;\Leftarrow \; |Y_a|=0.
$$

Finally, to verify (iii), notice that if $\mathcal{H}$ is radially unbounded in $x$, then $\mathcal{H}_{cl}$ and $\calw$ are radially unbounded in $w$, which implies that the stability properties of the equilibrium are global.\\
\end{IEEEproof}

\subsection{The case of unmatched disturbances} \label{subsec52}
{Our attention is now turned to considering the dynamics \eqref{iacl} with unmatched disturbances. That is, $d_a = 0_{m\times 1}$ and $d_u$ is non-zero. The approach taken here is to utilize Assumption \ref{Assumption:J12R12Const} to transform the dynamics into a similar system subject to a matched disturbance. Stability analysis of the transformed system then follows in much the same way as the matched disturbance case in Proposition \ref{propmatched}.

In order to transform the unmatched disturbance problem into a similar matched disturbance problem, first consider the closed-loop dynamics with the unmatched disturbance satisfying Assumption \ref{Assumption:J12R12Const} and $R_{c_2} = 0_{m\times m}$.
	\begin{equation}\label{UnmatchedDyn}
			\dot w
			=  
			(J_{cl} - R_{cl}) \nabla \mathcal{H}_{cl}
			-
			\begin{bmatrix}
				0_{m\times m} \\ (J_{au} + R_{au})^\top \\ 0_{m\times m}
			\end{bmatrix}
			\bar d_u.
	\end{equation}
The term $\bar d_u$ can be ``reflected" through the (2,1) block of $J_{cl} - R_{cl}$ where it then appears alongside the first element of $\nabla\mathcal{H}_{cl}$. However, this process results in an additional term which enters the system as a matched disturbance of the form described by Assumption \ref{Assumption:MatchedDist}. The transformed dynamics are given by
\begin{equation}
	\dot w
	=  
	( J_{cl}- R_{cl})
	\begin{bmatrix}
		\nabla_{w_a}\calH_{cl} + \bar d_u \\
		\nabla_{w_u} \calH_{cl} \\
		\nabla_{w_c} \calH_{cl}
	\end{bmatrix}
	-
	\begin{bmatrix}
		J_{c_1} - R_{c_1} \\ 0_{s\times m} \\ J_{c_1} - R_{c_1}
	\end{bmatrix}
	\bar d_u,
\end{equation}
By defining
\begin{equation}\label{bHcl}
 \mathbf{H}_{cl}(w)= \mathbf{H}(w_a,w_u) +  \hal\|w_c\|^2_{K_i},
\end{equation}
with $\mathbf{H}$ given in \eqref{bfh}, the dynamics are of the form
\begin{equation}\label{matcheDynamics}
	\dot w
	=  
	(J_{cl}- R_{cl})
	\nabla\mathbf{H}_{cl}
	-
	\begin{bmatrix}
		J_{c_1} - R_{c_1} \\ 0_{s\times m} \\ J_{c_1} - R_{c_1}
	\end{bmatrix}
	\bar d_u.
\end{equation}

\begin{proposition}\label{prounpmatched}
	Consider the system \eqref{phdist} with $d_a=0_{m\times 1}$, $d_u$ satisfying Assumption \ref{Assumption:J12R12Const} and the open-loop system \eqref{phdist} satisfying Assumption \ref{properAssump2}. Let the IAC be given by \eqref{controlLaw2} with the controller parameter $R_{c_2} = 0_{m\times m}$. 
\begin{itemize}
\item[(i)] The $(n+m)$-dimensional vector
\begin{equation}\label{unmatchedEquilibrium}
	\bar w = (\bar x_a,\bar x_u,K_i^{-1}\bar d_u)
\end{equation}
is a stable equilibrium of the closed-loop system. 
\item[(ii)] If the signal
\begin{equation}\label{detectOutput}
	Y_u:= \nabla_{w_a}\mathbf{H} + K_i \tilde w_c
\end{equation}
is a detectable output, the equilibrium is asymptotically stable.
\item[(iii)] The stability properties are global if the Hamiltonian function $\mathbf{H}$ is radially unbounded. 
\end{itemize}
\end{proposition}

\begin{IEEEproof}
As the disturbance satisfies Assumption \ref{Assumption:J12R12Const} and $R_{c_2} = 0_{m\times m}$, the closed-loop dynamics can be written as in \eqref{matcheDynamics}. To verify that $\bar w$ is indeed an equilibrium point, consider the gradient of $\mathbf{H}_{cl}$:
\begin{equation}
	\nabla\mathbf{H}_{cl}
	=
	\begin{bmatrix}
		\nabla_{x}\mathbf{H}  \\
		K_i w_c
	\end{bmatrix}
\end{equation}
By Assumption \ref{properAssump2}, $\nabla\mathbf{H}(\bar x) = 0_{n\times 1}$. This leads to
\begin{equation}\label{eqGrad}
	\nabla \bar{\mathbf{H}}_{cl}
	=
	\begin{bmatrix}
	    0_{m\times 1} \\
		0_{s\times 1} \\
		K_i \bar w_c
	\end{bmatrix}.
\end{equation}
Substitution of the gradient \eqref{eqGrad} into \eqref{matcheDynamics} yields
$$
\dot w|_{w = \bar w}  = \begmat{\bar J_{c_1} - \bar R_{c_1} \\ 0_{s\times 1}\\ \bar J_{c_1} - \bar R_{c_1} }(K_i \bar w_c- \bar d_u)=0_{(n+m) \times 1},
$$
which verifies that $\bar w$ is an equilibrium.

The proof of stability follows from similar argument as Proposition \ref{propmatched}. The shifted Hamiltonian function \eqref{shiham} becomes
$$
\mathbf{W}(w) :=\mathbf{H}_{cl}(w_a,w_u) - \bar d_u^\top(w_c-\bar w_c)-\bar {\mathbf{H}}_{cl},
$$
which is clearly positive-definite, therefore qualifies as a Lyapunov candidate for the closed-loop system. The time derivative of $\mathbf{H}_{cl}$ verifies
\begalis{
		\dot{\mathbf{H}}_{cl}
		\leq&
		 \|\nabla_{w_a}\mathbf{H}_{cl}+K_i w_c\|_{(J_{c_1}-R_{c_1})}^2 \\
		&-(\nabla_{w_a}\mathbf{H}_{cl}+K_i w_c)^\top (J_{c_1}-R_{c_1})\bar d_u.
}
where we used the fact that $R_{uu}>0$ to get the inequality. Likewise, the time derivative of the term $\bar d_u^\top w_c$ satisfies
\begalis{
		\bar d_u^\top \dot w_c
		&=
		\bar d_u^\top (J_{c_1}-R_{c_1})(\nabla_{w_a}\mathbf{H}_{cl}+K_i w_c-\bar d_u).
}
Combining these two equations we complete the squares and get the bound 
\begin{equation}\label{LaypTimeDer}
		\dot{\mathbf{W}} \leq -\|\nabla_{w_a}\mathbf{H}_{cl}+K_i \tilde w_c \|_{R_{c_1}}^2 \leq 0,
\end{equation}
which proves claim (i).

The proof of claim (ii) is established with LaSalle's invariance principle and the following implication
$$
\dot{\mathbf{W}} = 0\;\Leftarrow \; |Y_u|=0.
$$

Finally, to verify (iii), notice that if $\mathbf{H}$ is radially unbounded in $x$, then $\mathbf{H}_{cl}$ and $\mathbf{W}$ are radially unbounded in $w$, which implies that the stability properties of the equilibrium are global.
\end{IEEEproof}

\subsection{The case of matched and unmatched disturbances} 
\label{subsec44}

To close this section, we note that although the cases of matched and unmatched disturbances have been treated separately, the controller is able to reject the effects of both simultaneously. Indeed, if Assumptions \ref{Assumption:MatchedDist}-\ref{properAssump2} are satisfied and the controller parameters are chosen such that $R_{c_2} = 0_{m\times m}$ and $J_{c_1}, R_{c_2}$ satisfy \eqref{Jc1Rc1}, then the closed-loop is stable. 

The proof of this claim follows from the same line of reasoning as in the unmatched disturbance case in subsection \ref{subsec52}. The unmatched disturbance, which satisfies Assumption \ref{Assumption:J12R12Const}, is again ``reflected" through the $(2,1)$ block of $J_{cl}-R_{cl}$ so that the term $\bar d_u$ appear alongside the first element of $\nabla\mathcal{H}_{cl}$. However, this process of moving the unmatched disturbance again creates a similar matched disturbance. Recalling that $J_{c_1}-R_{c_1}=G_d$, the resulting dynamics have the form
\begin{equation}
	\dot w
	=  
	( J_{cl}- R_{cl})
	\nabla\mathbf{H}_{cl}
	-
	\begin{bmatrix}
		G_d \\ 0_{s\times m} \\ G_d
	\end{bmatrix}
	(\bar d_a + \bar d_u),
\end{equation}
where $\mathbf{H}_{cl}$ is defined in \eqref{bHcl}.

\begin{proposition}\label{proupmixed}
	Consider the system \eqref{phdist} with $d_a$ satisfying Assumption \ref{Assumption:MatchedDist}, $d_u$ satisfying Assumption \ref{Assumption:J12R12Const} and the open-loop system \eqref{phdist} satisfying Assumption \ref{properAssump2}. Let the IAC be given by \eqref{controlLaw2} with the controller parameter  $R_{c_2} = 0_{m\times m}$. 
\begin{itemize}
\item[(i)]  The $(n+m)$-dimensional vector
\begin{equation}\label{mixedEquilibrium}
	\bar w = (\bar x_a,\bar x_u,K_i^{-1}\left[\bar d_a + \bar d_u\right])
\end{equation}
is a stable equilibrium of the closed-loop system. 
\item[(ii)] If the signal $Y_u$ defined in \eqref{detectOutput} is a detectable output, the equilibrium is asymptotically stable.
\item[(iii)] The stability properties are global if the Hamiltonian function $\mathbf{H}$ is radially unbounded. 
\end{itemize}
\end{proposition}
\begin{IEEEproof}
The proof follows from the same procedure as the proof of Proposition \ref{prounpmatched} and is omitted for brevity.
\end{IEEEproof}

\section{Application to mechanical systems}\label{Mech}

{In this section, the IAC \eqref{controlLaw2} is applied to robustify energy-shaping controlled underactuated mechanical systems of the form \eqref{mecdist} with respect to constant matched disturbances. The problem considered here has been previously considered in \cite{Romero2013a} and \cite{Donaire2016} (see \cite{Donaire2016} for the detailed explanation and motivation of the problem). For convenience, we repeat the dynamics \eqref{mecdist} here: 
\begin{equation}\label{mecdist2}
	\begin{split}
		\begin{bmatrix}
	         \dot{q} \\
	         \dot{\bp}
		\end{bmatrix}
		&=
		\begin{bmatrix}
	          0_{{\ell}\times {\ell}} & M^{-1}(q)\mathbf{M}_d(q) \\
	         -\mathbf{M}_d(q)M^{-1}(q) & \mathbf{J}_2(q,\bp)-R_d(q)
		\end{bmatrix}
		\nabla \mathbf{H}_d \\
		&\phantom{---}+
		\begin{bmatrix}
		    d_u \\
		    G(q)(u-d_a)
		\end{bmatrix} \\
		\mathbf{H}_d(q,\mathbf{p})
		&= 
		\frac 12 \|\bp\|^2_{\mathbf{M}_d^{-1}(q)} + V_d(q).
	\end{split}
\end{equation}}

\subsection{Problem formulation}\label{sec61}
{Consider the dynamics of an energy-shaping controlled mechanical system \eqref{mecdist2} subject to a constant matched disturbance. That is, $d_a = \bar d_m$ for some constant $\bar d_m\in\mathbb{R}^{m}$. 
Define mappings $\mathbf{u}: \rea^l \times \rea^l \times \rea^m \to \rea^m$ and $\mathbf{F}:\rea^l \times \rea^l \to \rea^m$ such that the  IAC 
\begin{equation}
	\begin{split}
		u &= \mathbf{u}(q,\mathbf{p},{x}_c) \\
		\dot{x}_c &= \mathbf{F}(q,\mathbf{p}),
	\end{split}
\end{equation}
where ${x}_c\in\mathbb{R}^m$ is the state of the controller, that ensures the closed-loop is a pH system with an (asymptotically) stable equilibrium at $(q,\bp,{x}_c) = (q^{\star},0,{x}_c^\star)$ for some ${x}_c^\star\in\mathbb{R}^m$.}
\subsection{Momentum transformation}
{The system \eqref{mecdist2} is similar to the system \eqref{phdist} while the problem formulation of subsection \ref{sec61} mirrors that of of subsection \ref{subsec23}. Thus, it makes sense to apply the IAC \eqref{controlLaw2} as a solution to the disturbed mechanical system problem.

The key difference between the systems \eqref{mecdist2} and \eqref{phdist} is that in \eqref{mecdist2}, the control input $u$ is pre-multiplied by a matrix $G(q)$. To compensate for this difference, we present a momentum transformation that allows the dynamics \eqref{mecdist2} to be expressed in a similar form where the input is pre-multiplied by the identity matrix. Slightly different forms of the following lemma have been used in the literature (e.g. \cite[Lemma 2]{Fujimoto2001a}, \cite[Proposition 1]{Venkatraman2010a} and \cite[Theorem 1]{Duindam208}).}

\begin{lemma}\label{momLemma}
	Consider the system \eqref{mecdist2} under the change of momentum
	\begin{equation}\label{momtransf}
		(q,p)  = (q,T(q)\mathbf{p}).
	\end{equation}
	where $T(q)\in\mathbb{R}^{l\times l}$ is invertible.
Then, the dynamics \eqref{mecdist2} can be equivalently expressed as
\begin{equation}\label{momTrans}
			\begin{bmatrix}
		         \dot{q} \\
		         \dot{p}
			\end{bmatrix}
			=
			\begin{bmatrix}
		          0_{l\times l} & Q \\
		         -Q^\top & C-D
			\end{bmatrix}
			\begin{bmatrix}
				\nabla_q \calH \\ \nabla_p \calH
			\end{bmatrix} 
			+
			\begin{bmatrix}
			    0_{l\times m} \\
			    TG
			\end{bmatrix}
			(u-\bar d_m),
\end{equation}
with
\begin{equation}\label{HMech}
\calH(q,p):=\frac12 p^\top M_d^{-1}(q)\;p + V_d,
\end{equation}
and
\begin{equation}\label{Fdefn}
		\begin{split}
			M_d^{-1}(q)&=T^{-\top}\; \mathbf M_d^{-1}\; T^{-1}, \\
			Q(q) &= M^{-1}\; \mathbf{M}_d \; T^\top, \\
			C(q,p) &= \left[\nabla_q^\top(T\mathbf{p}) \, M^{-1}\mathbf{M}_d - \mathbf{M}_d M^{-1} \, \nabla_q(T\mathbf{p}) \right.\\
				&\left. \phantom{--}+ T \mathbf \,  \mathbf{J}_u \, T^\top \right] \big|_{\mathbf{p} = T^{-1}(q)p}  \\
			D(q,p) &=  T \; R_d \; T^\top  \big|_{\mathbf{p} = T^{-1}p}.
		\end{split}
	\end{equation}
\end{lemma}

\begin{IEEEproof} The proof of this lemma follows the same as the proof in \cite[Lemma 2]{Fujimoto2001a}, \cite[Proposition 1]{Venkatraman2010a} and \cite[Theorem 1]{Duindam208}, therefore the full proof is omitted.
\end{IEEEproof}

Consider now the transformation matrix $T(q)$ defined as
\begin{equation}\label{Ttransform}
	T(q)=
	\begin{bmatrix}
		\{G(q)^\top G(q)\}^{-1}G^\top(q) \\ G^\perp(q)
	\end{bmatrix},
\end{equation}
where $G^\perp(q)$ is a full-rank left annihilator of $G(q)$. It follows that
\begin{equation}
		T(q)G(q) = \begin{bmatrix} I_{m} \\ 0_{r\times m} \end{bmatrix},
\end{equation}
where $r := l-m$. Considering the momentum vector in \eqref{momTrans} as $p = \operatorname{col}(p_a,p_u)$, where $p_a\in\mathbb{R}^m$ and $p_u\in\mathbb{R}^{r}$, and the matrices $C, D$ and $Q$ as
\begin{equation}
	\begin{split}
		C
		=
		\begin{bmatrix}
			C_{aa} & C_{au} \\
			-C_{au}^\top & C_{uu}
		\end{bmatrix}, \
		D
		=
		\begin{bmatrix}
			D_{aa} & D_{au} \\
			D_{au}^\top & D_{uu}
		\end{bmatrix}, \
		Q
		=
		\begin{bmatrix}
			Q_a & Q_u
		\end{bmatrix},
	\end{split}
\end{equation}
where $C_{aa}, D_{aa} \in \mathbb{R}^{m\times m}$, $C_{au}, D_{au} \in \mathbb{R}^{m\times r}$, $C_{uu}, D_{uu} \in \mathbb{R}^{r\times r}$, $Q_a\in\mathbb{R}^{l\times m}$, $Q_u\in\mathbb{R}^{l\times r}$, the system \eqref{mecdist2}, under the change of coordinates \eqref{momtransf} where $T(q)$ is defined by \eqref{Ttransform}, can be expressed as
\begin{equation}\label{mecp}
	\begin{split}
		\begin{bmatrix}
       \dot{p}_a \\
       \dot{p}_u \\
       \dot{q}
		\end{bmatrix}
		&=
		\begin{bmatrix}
        C_{aa}-D_{aa} & C_{au}-D_{au} & -Q_a^\top \\
        -C_{au}^\top-D_{au}^\top & C_{uu}-D_{uu} & -Q_u^\top \\
       Q_a & Q_u & 0_{l\times l}
		\end{bmatrix}
		\begin{bmatrix}
			\nabla_{p_a} \calH \\ \nabla_{p_u} \calH \\ \nabla_q \calH
		\end{bmatrix} \\
		&\phantom{---}+
		\begin{bmatrix}
			I_{m} \\ 0_{r\times m} \\ 0_{l\times m}
		\end{bmatrix}
		(u-\bar d_m).
	\end{split}
\end{equation}
Notice that the system \eqref{mecp} is in the form \eqref{phdist} with $x_a = p_a$, $x_u = \text{col}(p_u,q)$, 
	\begin{equation*}
		\begin{split}
		J_{aa}(x_a,x_u) &= C_{aa} \\
		J_{au}(x_a,x_u) &= \begin{bmatrix} C_{au} & -Q_a^\top \end{bmatrix} \\
		J_{uu}(x_a,x_u) &= 
		\begin{bmatrix}
			C_{uu} & -Q_u^\top \\
			Q_u & 0_{l\times l}
		\end{bmatrix} \\
		\end{split}
	\end{equation*}
	\begin{equation}\label{MechMatrixDefs}
		\begin{split}
		R_{aa}(x_a,x_u) &= D_{aa} \\
		R_{au}(x_a,x_u) &= \begin{bmatrix} D_{au} & 0_{m\times l} \end{bmatrix} \\
		R_{uu}(x_a,x_u) &= 
		\begin{bmatrix}
			D_{uu} & 0_{r\times l} \\
			0_{l\times r} & 0_{l\times l}
		\end{bmatrix}.
		\end{split}
	\end{equation}
\subsection{Integral action controller}
{
In the previous section it was shown that the system \eqref{mecdist2} can be equivalently written as \eqref{mecp}, which falls into the class of systems \eqref{phdist}. Thus, we can now utilise the control law \eqref{controlLaw2} to solve the disturbance rejection problem. The following proposition formalises the stability properties of the closed-loop.

 \begin{proposition}\label{MechInt} 
Consider the dynamics \eqref{mecp}, or equivalently \eqref{mecdist2}, in closed-loop with the controller 
\begin{eqnarray}
	\hspace{-4mm}   u \hspace{-2mm}&=&\hspace{-2mm} (-C_{aa}+D_{aa}+J_{c_1}-R_{c_1}-R_{c_2})\nabla_{p_a} \calH \nonumber \\ 
	\hspace{-4mm}   && \hspace{-4mm}+ (J_{c_1}-R_{c_1}) K_i (p_a-{x}_c) + 2 D_{au} \nabla_{p_u} \calH  \nonumber \\
	\hspace{-4mm}   \dot{x}_c\hspace{-2mm}&=&\hspace{-2mm}-R_{c_2}\nabla_{p_a}\mathcal{H} + (C_{au} + D_{au}) \nabla_{p_u}\mathcal{H} - Q_a^\top\nabla_q \mathcal{H} \label{controlLawMech2}
\end{eqnarray}
where $J_{c_1}$, $R_{c_1}$, $R_{c_2}, K_i\in\mathbb{R}^{m\times m}$ are constant tuning parameters chosen to satisfy $J_{c_1}=-J_{c_1}^\top$ and $R_{c_1}$, $R_{c_2}, K_i > 0$. 
\begin{itemize}
\item[(i)] 
The $(2l+m)$-dimensional vector
\begin{equation}\label{mechEq}
	(q,p,w_c)=(q^\star,0_{l\times 1},K_i^{-1}(J_{c_1}-R_{c_1})^{-1}\bar d_m).
\end{equation}
is a stable equilibrium of the closed-loop system.
\item[(ii)]
If the output
\begin{equation}
Y_{m} := \begin{bmatrix} \nabla_{p_a}\mathcal{H} \\ \tilde w_c \end{bmatrix} = \begin{bmatrix} \nabla_{p_a}\mathcal{H} \\ p_a-{x}_c+K_i^{-1}(J_{c_1}-R_{c_1})^{-1}\bar d_m \end{bmatrix}
\end{equation}
is detectable, then the equilibrium is asymptotically stable.
\item[(iii)]
The stability results are global if $V_d$ is radially unbounded.
\end{itemize} 
 \end{proposition}
 \begin{IEEEproof} The proof follows from direct application of Propositions \ref{propiacl} and \ref{propmatched}. As \eqref{mecp} is a matched disturbance problem, it must be verified that the disturbances satisfy Assumption \ref{Assumption:MatchedDist}. This can be seen to be true by taking $G_d = J_{c_1}-R_{c_1}$ for any constant $J_{c_1},R_{c_1}$. Then, the disturbance can be written as
 \begin{equation}
	 \bar d_m = \underbrace{(J_{c_1}-R_{c_1})}_{G_d}\underbrace{(J_{c_1}-R_{c_1})^{-1}\bar d_m}_{\bar d_a},
 \end{equation}
 verifying Assumption \ref{Assumption:MatchedDist}.
 
 As the system \eqref{mecp} in the form \eqref{phdist} and the controller \eqref{controlLawMech2} has the form \eqref{controlLaw2}, then, by Proposition \ref{propiacl}, the closed-loop dynamics can be written in the form \eqref{iacl}.
 
 Claims (i) and (ii) then follow from direct application of Proposition \ref{propmatched}. Also by Proposition \ref{propmatched}, (iii) is true if $\mathcal{H}_d$ is radially unbounded. Considering the definition of $\mathcal{H}_d$ in \eqref{HMech} and noting that $M_d > 0$, $\mathcal{H}_d$ is radially unbounded if $V_d$ is radially unbounded as desired.
\end{IEEEproof}}

The result in Proposition \ref{MechInt} can be tailored to fully-actuated mechanical systems, and since detectability can be easily shown in that case, then asymptotic stability of the desired equilibrium $(q^\star,p^\star,{x}_c^\star)$ is ensured.

%
\section{Examples}\label{ex}
{In this section, the proposed IAC is implemented on three examples. First, the IAC is applied to the PMSM with unknown load torque and unknown mechanical friction. Interestingly, in this example, the IAC can be implemented without knowledge of the motors angular velocity. The second example is a 2-degree of freedom (DOF) manipulator with unknown damping subject to a matched disturbance. The final example is the vertical take-off and landing (VTOL) system, an example of an underactuated mechanical system subject to a matched disturbance.}
\subsection{PMSM with mechanical friction and unknown load torque}
The PMSM is described by the dynamics \cite{Petrovic2001}:
\begin{equation}\label{PMSM}
	\begin{split}
		L_d\dot i_d
		&=
		-R_si_d + \omega L_qi_q + v_d \\
		L_q\dot i_q
		&=
		-R_si_q - \omega L_di_d - \omega\Phi + v_q \\
		J\dot{\omega}
		&=
		n_p
		\left[
		(L_d-L_q)i_di_q + \Phi i_q
		\right] - R_m\omega - \tau_L
	\end{split}
\end{equation}
where $i_d, i_q$ are currents, $v_d,v_q$ are voltage inputs, $n_p$ is the number of pole pairs, $L_d, L_q$ are the stator inductances, $\Phi$ is the back emf constant, $J$ is the moment of inertia, $R_s$ is the electrical resistance, $R_m$ is the mechanical friction coefficient and $\tau_L$ is a constant load torque.

Using the energy shaping controller proposed in \cite{Ortega2004}, the closed-loop has an asymptotically stable equilibrium at $(0,0,\omega^\star)$ when $\tau_L = 0$ and $R_m = 0$. The closed-loop dynamics have the pH representation
\begin{equation}\label{CLsys}
	\begin{split}
		\begin{bmatrix}
			\dot i_q \\ \dot i_d \\ \dot{\omega}
		\end{bmatrix}
		&=
		\begin{bmatrix}
			-r_2 & -C_{12}(i_d,i_q) & C_{23} \\
			C_{12}(i_d,i_q) & -r_1 & C_{13}(i_q) \\
			-C_{23} & -C_{13}(i_q) & -\frac{R_m}{J\gamma_2}
		\end{bmatrix}
		\begin{bmatrix}
			\nabla_{i_q}\calh \\ \nabla_{i_d}\calh \\ \nabla_{\omega}\calh
		\end{bmatrix} \\
		&\phantom{---}+
		\begin{bmatrix}
			u \\ 0 \\ 0
		\end{bmatrix}
		-
		\begin{bmatrix}
			0 \\ 0 \\ \frac{1}{J}\left(\tau_L + R_m\omega^\star\right)
		\end{bmatrix},
	\end{split}
\end{equation}
where
\begin{equation}
	\calh(i_d,i_q,\omega)
	=
	\frac12\gamma_1i_d^2 - \frac{1}{2C_{23}}\frac{n_p}{J}\Phi i_q^2 + \frac12\gamma_2(\omega-\omega^\star)^2,
\end{equation}
$r_1,r_2,\gamma_1,\gamma_2,C_{23}\in\mathbb{R}$ are tuning parameters satisfying $r_1,r_2,\gamma_1,\gamma_2 >0$, $C_{23} <0$, $C_{12}:\rea^2 \to \rea$ is a free function,
$$
C_{13}(i_q):= -\frac{n_p}{J\gamma_1}(L_d-L_q)i_q,
$$
and $u\in\mathbb{R}$ is an additional voltage input for IAC design. 

Taking $x_a = i_q$, $x_u = \operatorname{col}(i_d,\omega)$, the system is of the form \eqref{phdist} with
\begin{multicols}{2}
	\noindent
	\begin{equation*}
		\begin{split}
		J_{aa} &= 0 \\
		J_{au} &= \begin{bmatrix} -C_{12}(i_d,i_q) & C_{23} \end{bmatrix} \\
		J_{uu} &= \begin{bmatrix} 0 & C_{13}(i_q) \\ -C_{13}(i_q) & 0 \end{bmatrix} \\
		d_u &= \operatorname{col}\left(0,\frac{1}{J}\left[\tau_L + R_m\omega^\star\right]\right).
		\end{split}
	\end{equation*}
	\begin{equation}\label{DisRej:ISOPHS:11}
		\begin{split}
		R_{aa} &= r_2 \\
		R_{au} &= 0_{1\times 2} \\
		R_{uu} &= \operatorname{diag}\left(r_1, \frac{R_m}{J\gamma_2}\right)
		\end{split}
	\end{equation}
\end{multicols}
\noindent
Notice that the disturbance is unmatched. Our objective is now to apply the IAC \eqref{controlLaw2} to guarantee that the disturbed system has a stable equilibrium satisfying $\nabla_{x_u}\calh = 0_{2\times 1}$. That is, the disturbed equilibrium should satisfy $\bar \omega = \omega^\star$ and $\bar i_d = 0$.
Before integral action can be applied, it must be verified that \eqref{CLsys} satisfies Assumptions \ref{Assumption:J12R12Const} and \ref{properAssump2}.

\begite
\item[\bf{A\ref{Assumption:J12R12Const}.}]
For the system to satisfy Assumption \ref{Assumption:J12R12Const}, the unmatched disturbance should be written as
\begin{equation}
	d_u
	=
	\begin{bmatrix}
		0 \\ \frac{1}{J}\left(\tau_L + R_m\omega^\star\right)
	\end{bmatrix}
	=
	\begin{bmatrix} -C_{12}(i_d,i_q) \\ C_{23} \end{bmatrix}\bar d_u
\end{equation}
for some constant $\bar d_u$. This is only possible if $C_{12} = 0$. As this is a free function, we make this selection for the energy shaping controller. With this choice,
\begin{equation}
	\bar d_u = \frac{1}{JC_{23}}\left(\tau_L + R_m\omega^\star\right).
\end{equation}

\item[\bf{A\ref{properAssump2}.}]
The shifted Hamiltonian \eqref{bfh} takes the form
\begin{equation}
	\begin{split}
		\mathbf{H} 
		&= 
		\frac12\gamma_1i_d^2 - \frac{1}{2C_{23}}\frac{n_p}{J}\Phi i_q^2 + \frac12\gamma_2(\omega-\omega^\star)^2 \\ 
		&\phantom{---}+ i_q\frac{1}{JC_{23}}\left(\tau_L + R_m\omega^\star\right) \\
		&= 
		\frac12\gamma_1i_d^2 - \frac{1}{2C_{23}}\frac{n_p}{J}\Phi \left(i_q-\frac{\tau_L+R_m\omega^\star}{n_p\Phi}\right)^2 \\ 
		&\phantom{---}+ \frac12\gamma_2(\omega-\omega^\star)^2 + (\tau_L+R_m\omega^\star)^2\frac{1}{2n_pJC_{23}\Phi} \\
	\end{split}
\end{equation}
which is clearly minimised at the desired point
\begin{equation}\label{eqpoint}
	(\bar i_q, \bar i_d, \bar \omega)
	=
	\left(\frac{\tau_L+R_m\omega^\star}{n_p\Phi}, 0, \omega^\star\right).
\end{equation}
\endite

As the system satisfies Assumptions \ref{Assumption:J12R12Const} and \ref{properAssump2}, integral action can be applied using the control law \eqref{controlLaw2}. Taking $R_{c_1} = R_{aa}$, $J_{c_1} = 0$, $R_{c_2} = 0$, the control law simplifies to
\begin{equation}\label{PMSM:controlLaw}
	\begin{split}
		u &= -r_2K_i(i_q - {x}_c) \\
		\dot{x}_c&= C_{23} \nabla_{\omega}\calh.
	\end{split}
\end{equation}
Since the system \eqref{CLsys} is subject to an unmatched disturbance only, the IAC can be implemented in the $w_c$ coordinates as per \eqref{controlLaw3}. Using this realisation, the controller simplifies to be
\begin{equation}\label{PMSM:CbiCtrl}
	\begin{split}
		u &= -r_2K_i w_c \\
		\dot{w}_c &= -r_2(\nabla_{i_q}H_d+K_i w_c).
	\end{split}
\end{equation}
where $w_c\in\mathbb{R}$ is the state of the controller. To reiterate, the expressions \eqref{PMSM:controlLaw} and \eqref{PMSM:CbiCtrl} describe the same IAC but are expressed in different coordinates. Interestingly, the IAC \eqref{PMSM:CbiCtrl} is independent of the rotor speed $\omega$ whereas the realisation \eqref{PMSM:controlLaw} is not.

{By Proposition \ref{prounpmatched}, the point \eqref{eqpoint} is a stable equilibrium of the closed-loop system. Further, the system is asymptotically stable if the output
\begin{equation}
	\begin{split}
		Y_u&= \nabla_{w_a}\mathbf{H} + K_i \tilde w_c \\
		&=
		\nabla_{i_q}\calh + \bar d_u + K_i(w_c-K_i^{-1}\bar d_u) \\
		&=
		\nabla_{i_q}\calh + K_iw_c
	\end{split}
\end{equation}
is detectable. To verify this fact, we set $Y_u = 0$ identically and investigate whether this implies that $|w(t)| \to 0$, with $w=(i_d,i_q,\omega,w_c)$. Defining the set $\mathcal{Y}_u=\{w\in\mathbb{R}^4|Y_u=0\}$, the dynamics of $w_c$ restricted to $\mathcal{Y}_u$ satisfy $\dot w_c|_{\mathcal{Y}_u=0} = 0$ which implies that $w_c|_{\mathcal{Y}_u=0}$ is constant. By the dynamics of $w_c$ in \eqref{PMSM:CbiCtrl}, $i_q|_{\mathcal{Y}_u=0}$ must be constant and $\dot i_q|_{\mathcal{Y}_u=0} = 0$. Considering the dynamics of $i_q$ in \eqref{CLsys} and recognising that $u|_{\mathcal{Y}_u=0} = -r_2K_iw_c = r_2\nabla_{i_q}H_d$, $\dot i_q|_{\mathcal{Y}_u=0} = 0$ implies that $\omega|_{\mathcal{Y}_u=0}=\omega^\star$ and $\dot{\omega}|_{\mathcal{Y}_u=0}=0$. The dynamics of $i_d$ in \eqref{CLsys} now reduce to $\dot i_d|_{\mathcal{Y}_u=0}=-r_1\nabla_{i_d}H_d$ which implied that $i_d,\dot i_d$ tend to zero. To complete the argument, consider the dynamics of $\omega$ in \eqref{CLsys} which implies that $-C_{23}\nabla_{i_q}H_d|_{\mathcal{Y}_u=0} \to \frac{1}{J}\left(\tau_L + R_m\omega^\star\right)$, which can be used to recover the equilibrium \eqref{eqpoint}.}

\subsection{2-degree of freedom manipulator with unknown damping}
The 2-degree of freedom (DOF) planar manipulator in closed-loop with energy shaping control has the dynamic equations
\begin{equation}
	\begin{split}
		\begin{bmatrix}
			\dot q \\ \dot p
		\end{bmatrix}
		&=
		\begin{bmatrix}
			0_{2\times 2} & I_{2\times 2} \\
			-I_{2\times 2} & -R_d
		\end{bmatrix}
		\begin{bmatrix}
			\nabla_q \mathcal{H} \\
			\nabla_p \mathcal{H}
		\end{bmatrix}
		+
		\begin{bmatrix}
			0_{2\times 2} \\
			I_{2\times 2}
		\end{bmatrix}
		(u-\bar d_a) \\
		y
		&=
		\nabla_p \mathcal{H} \\
		\mathcal{H}
		&=
		\frac12 p^\top M^{-1}(q)p + \frac12(q-q^\star)^\top K_p(q-q^\star)
	\end{split}
\end{equation}
where
\begin{equation}
	M(q)
	=
	\begin{bmatrix}
		a_a+a_u+2b\cos\theta_u & a_u+b\cos\theta_u \\
		a_u+b\cos\theta_u & a_u
	\end{bmatrix},
\end{equation}
$q = (\theta_a,\theta_u)$, $p = M(q)\dot q$, $a_a, a_u, b$ are constant system parameters, $K_p\in \mathbb{R}^{2\times 2}$ is positive definite and $\bar d_a\in\mathbb{R}^2$ is a constant disturbance \cite{Dirksz2012a}. $R_d\in \mathbb{R}^{2\times 2}$ is an unknown, positive definite matrix arising from physical damping. Taking $x_1 = p, x_2 = q$, this system is of the form \eqref{phdist} with
\begin{multicols}{2}
	\noindent
	\begin{equation*}
		\begin{split}
		J_{aa} &= 0_{2\times 2} \\
		J_{au} &= -I_{2\times 2} \\
		J_{uu} &= 0_{2\times 2} \\
		\end{split}
	\end{equation*}
	\begin{equation}
		\begin{split}
		R_{aa} &= R_d \\
		R_{au} &= 0_{2\times 2} \\
		R_{uu} &= 0_{2\times 2}.
		\end{split}
	\end{equation}
\end{multicols}
As discussed in R8. of subsection \ref{subsec42}, as $R_{aa}$ is constant and $R_{au} = 0_{2\times 2}$, the IAC can be applied to the system without knowledge of the damping parameter $R_d$. The simplified IAC is given by
\begin{eqnarray} 
		u &=& -R_{c_2}(x)M^{-1}(q)p -\kappa(p-{x}_c) \nonumber \\
		\dot{x}_c&=&-R_{c_2}(x)M^{-1}(q)p - \nabla_q \mathcal{H},
\end{eqnarray}
where $R_{c_2}>0$ and $\kappa>0$ are free to be chosen. 

The simplicity of this IAC should be compared with the ones proposed in \cite{Romero2013a} to reject constant matched disturbances that, moreover, are not applicable in the presence of unknown damping.  

\subsection{Damped VTOL aircraft}
The dynamics of the VTOL aircraft with physical damping in closed loop with the IDA-PBC law proposed in \cite{Acosta2005,Gomez-Estern2004} are of the form
\begin{eqnarray}\label{Mech:VTOLEngShap}
		 \begin{bmatrix}
			\dot{q} \\ \dot{\mathbf{p}}
		\end{bmatrix}
		\hspace{-2mm}&=&\hspace{-2mm}
		\begin{bmatrix}
			0_{3\times 3} & M^{-1}\mathbf{M}_d \\
			-\mathbf{M}_dM^{-1} & \mathbf{J}_u-GK_vG^\top-R_0\mathbf{M}_d
		\end{bmatrix}
		\begin{bmatrix}
			\nabla_q \mathbf{H}_d \\ \nabla_\mathbf{p} \mathbf{H}_d
		\end{bmatrix} \nonumber  \\
		&&\phantom{-}+
		\begin{bmatrix}
			0_{3\times 2} \\ G(q)
		\end{bmatrix}
		(u-\bar d_m) \nonumber \\
		\mathbf{H}_d&=&\frac12 \|\bp\|^2_{\mathbf{M}_d^{-1}(q)} + V_d(q),
\end{eqnarray}
where $q = (x, y, \theta)$ with $x,y$ denoting the translational position and $\theta$ the angular position of the aircraft, $\mathbf{p} = \dot q$ is the momentum,
\begin{equation}
	\begin{split}
		V_d(q)
		&=
		\frac{g(1-\cos \theta)}{k_1 - k_2\epsilon} + \frac{1}{2}[z(q) - z(q^\star)]^\top P[z(q) - z(q^\star)] \\
		z(q)
		&=
		\begin{bmatrix}
			x - x^* - \frac{k_3}{k_1 - k_2\epsilon}\sin \theta \\
			y - y^* - \frac{k_3 - k_1\epsilon}{k_1 - k_2\epsilon}(\cos \theta - 1)
		\end{bmatrix} \\
		\mathbf{M}_d(q)
		&=
		\begin{bmatrix}
			k_1\epsilon\cos^2 \theta + k_3 & k_1\epsilon\cos \theta\sin \theta & k_1\cos \theta \\
			k_1\epsilon\cos \theta\sin \theta & -k_1\epsilon\cos^2 \theta + k_3 & k_1\sin \theta \\
			k_1\cos \theta & k_1\sin \theta & k_2
		\end{bmatrix} \\
		R_0(q)
		&=
		\operatorname{diag}(r_1(q),r_2(q),r_3(q)) \\
		\mathbf{J}_2(q,\tilde p)
		&=
		\begin{bmatrix}
			0 & \tilde{p}^\top\alpha_1(q) & \tilde{p}^\top\alpha_2 \\
			-\tilde{p}^\top\alpha_1(q) & 0 & \tilde{p}^\top\alpha_3 \\
			-\tilde{p}^\top\alpha_2 & -\tilde{p}^\top\alpha_3 & 0
		\end{bmatrix} \\
		\tilde{p}(q,\mathbf{p}) &= \mathbf{M}_d^{-1}(q)\mathbf{p} \\
		\alpha_1(q)
		&=
		-\frac12 k_1\gamma_{30}
		\begin{bmatrix}
			2\epsilon\cos \theta & 2\epsilon\sin \theta & 1
		\end{bmatrix}^\top \\
		\alpha_2
		&=
		-\frac12 k_1\gamma_{30}
		\begin{bmatrix}
			0 & 1 & 0
		\end{bmatrix}^\top \\
		\alpha_3
		&=
		-\frac12 k_1\gamma_{30}
		\begin{bmatrix}
			-1 & 0 & 0
		\end{bmatrix}^\top \\
		\gamma_{30} &= k_1 - \epsilon k_2 \\
		G(q)
		&=
		\begin{bmatrix}
			1 & 0 \\
			0 & 1 \\
			\frac{1}{\epsilon}\cos{\theta} & \frac{1}{\epsilon}\sin{\theta}
		\end{bmatrix},
	\end{split}
\end{equation}
$g$ is the acceleration due to gravity, $\epsilon$ is a constant that describes the coupling effect between the translational and rotational dynamics, $r_i(q) > 0$ are the damping coefficients, $\bar d_m\in \mathbb{R}^2$ is a constant disturbance and $u\in \mathbb{R}^2$ is a control input for additional control design. The parameters $K_v,P\in\mathbb{R}^{2\times 2}$ and $k_1, k_2, k_3 \in \mathbb{R}$ are tuning parameters of the IDA-PBC law that should be chosen to satisfy the conditions given in Section 7.1 of \cite{Gomez-Estern2004} to ensure stability of the closed-loop.

In order to enhance the robustness of the control system with an integral action controller, we perform the momentum transformation \eqref{momtransf} with $T(q)$ defined in \eqref{Ttransform}, which yields
\begin{equation}
	T(q)
	=
	\begin{bmatrix}
		\frac{\epsilon^2 + \sin^2 q_3}{\epsilon^2 + 1} & -\frac{\sin(2q_3)}{2(\epsilon^2 + 1)} & \frac{\epsilon\cos(q_3)}{\epsilon^2 + 1} \\
		-\frac{\sin(2q_3)}{2(\epsilon^2 + 1)} & \frac{\epsilon^2 - \sin^2 q_3 + 1}{\epsilon^2 + 1} & \frac{\epsilon\sin q_3}{\epsilon^2 + 1} \\
		\cos(q_3) & \sin(q_3) & -\epsilon
	\end{bmatrix}.
\end{equation}
Then, the dynamics \eqref{Mech:VTOLEngShap} written in coordinates $(q,p)$ take the form \eqref{mecp}. The integral action control law \eqref{controlLawMech2} can be applied and the closed-loop system is stable by Proposition \ref{MechInt}. Since $\mathbf{H}_d$ is positive definite and proper, as discussed in \cite{Acosta2005,Gomez-Estern2004}, the dynamics \eqref{Mech:VTOLEngShap} in closed-loop with the integral controller \eqref{controlLawMech2} ensures that the desired equilibrium  $(q,\mathbf{p},{x}_c) = (q^*,0_{3\times 1},K_i^{-1}\bar d_m)$ is globally stable. Furthermore, noting that $R_0 > 0$, $p(t)\to 0$. Combining this fact with the detectability arguments contained within Proposition 8 of \cite{Acosta2005}, the closed-loop is almost globally asymptotically stable.

\subsection{Damped VTOL aircraft simulation}
The energy-shaping controlled VTOL system subject to constant disturbance in closed-loop with the IAC \eqref{controlLawMech2} was numerically simulated under the following scenario: the initial conditions are $q(0) = (-5,0,0.1)$, $\mathbf{p}(0) = (-0.1,-0.1,0.1)$ and ${x}_c=(0,0)$, the desired configuration is $q^* = (5,0,0)$ and $\mathbf{p}^* = (0,0,0)$. The values of the model parameters and controller gains are $R_0 = I_{3\times 3}$, $\epsilon = 1$, $k_1 = 2$, $k_2 = 1.1$, $k_3 = 30$, $J_{c_1} = 0_{2\times 2}$, $K_i = I_{2\times 2}$
$$
K_v = \begin{bmatrix} 10 & 5 \\ 5 & 10 \end{bmatrix},\; P = \begin{bmatrix} 0.03 & 0 \\ 0 & 0.02 \end{bmatrix},
$$
$$
R_{c_1} = \begin{bmatrix} 10 & 5 \\ 5 & 10 \end{bmatrix},\;
R_{c_2} = \begin{bmatrix} 10 & 0 \\ 0 & 10 \end{bmatrix}.
$$

Figure \ref{VTOLfig2} shows the time histories of the configuration variables, the momentum, the controller states and Lyapunov function. On the time interval $[0,30)$, the system operates without disturbance and tends towards the desired configuration, as expected. At $t=30s$, a matched disturbance $\bar d_m = (5,-5)$ was applied to the system. The IAC rejects the effect of the disturbance and causes the VTOL system to tend towards the desired configuration. 

\begin{figure}[htbp]
	\centering
	\includegraphics[width=0.49\textwidth]{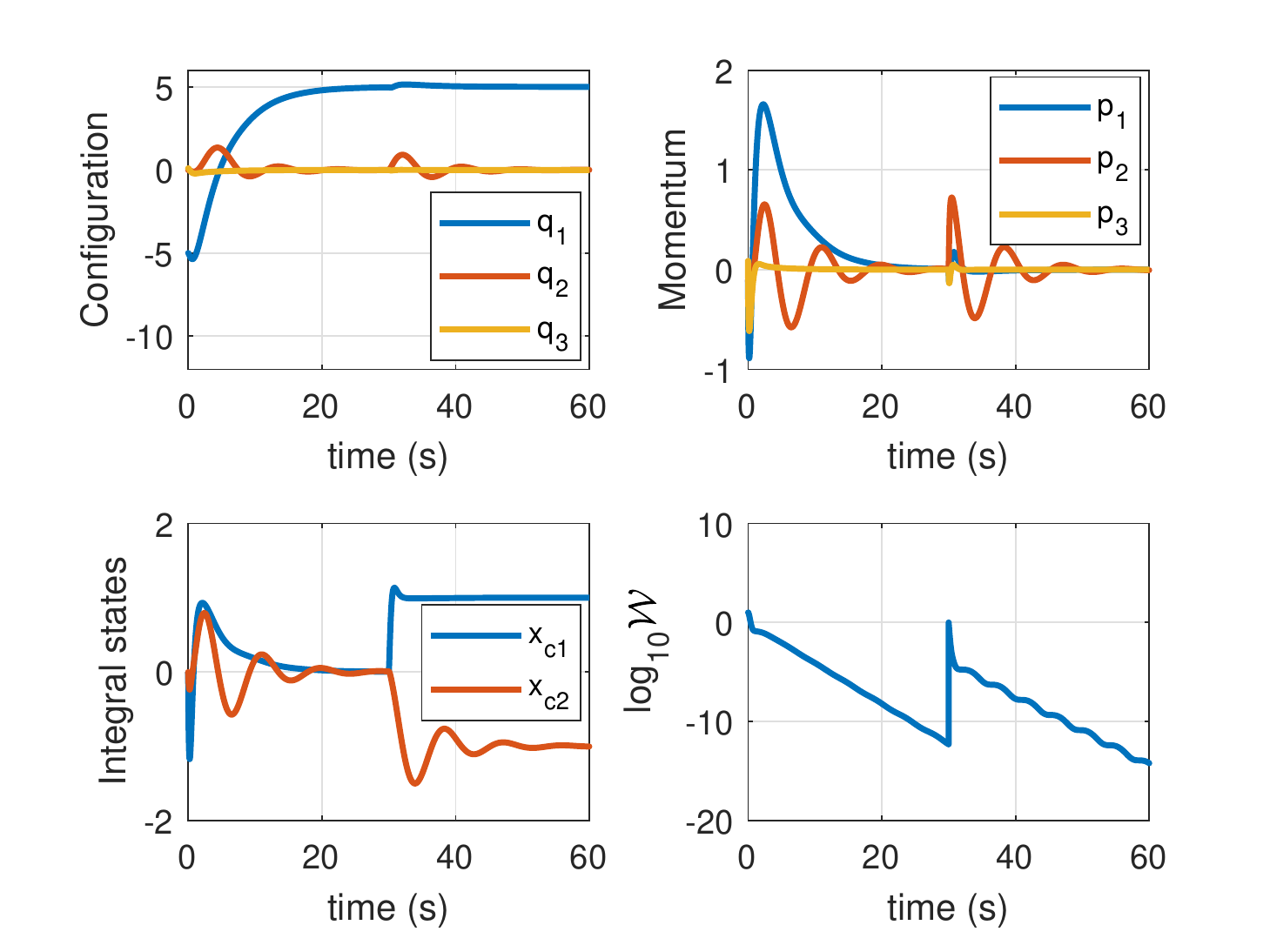}
	\caption{Simulation of the VTOL aircraft in closed-loop with an energy shaping plus integral action controller.}
	\label{VTOLfig2}
\end{figure}

\section{Conclusion}\label{con}
In this paper, a method for designing IAC for pH system subject to matched and unmatched disturbances is presented. The proposed design extends our previous work by relaxing the restrictive assumptions of a strongly convex and separable Hamiltonian function for the open-loop system. By relaxing these assumptions, the proposed IAC is shown to be applicable to a more general class of mechanical systems that strictly contains the class considered in previous works. The method is illustrated on three examples: a PMSM with unknown load torque and mechanical friction, 2-DOF manipulator with unknown damping and a VTOL aircraft.

\appendix\label{app1}

\begin{lemma}\label{AssumpRelation}
	Consider the system \eqref{phdist} and assume there exists some $\bar x$ verifying Assumption \ref{properAssump2}. If there exists a suitable coordinate transformation \eqref{zTranform}, then there exists some $\hat x$ satisfying \eqref{barxiacort}.
\end{lemma}
\begin{IEEEproof}
	By Assumption \ref{properAssump2}, there exists some $\bar x$ satisfying $\nabla_{x_a}\bar{\mathcal{H}}=-\bar d_u, \nabla_{x_u}\bar{\mathcal{H}}=0_{m\times 1}$. Using this point, we define $(\bar z_1, \bar z_2) = (\bar x_1, \bar x_2)$. Using this definition, the point $\hat x$ can be defined using the functions $\psi$ and its inverse $\mu$:
	\begin{equation}
		(\hat x_a, \hat x_u) = (\mu(\bar z_1, \bar z_2, \bar d_u), \bar z_2).
	\end{equation}
	It must now be verified that $\hat x$ satisfies \eqref{barxiacort}. To see that $\hat x\in\cale$, first notice that $\nabla_{x_a}\mathcal{H}|_{x_a=\hat x_1} = \nabla_{x_a}\mathcal{H}|_{x_a=\bar z_1}=-\bar d_u$ and  $\nabla_{x_u}\mathcal{H}|_{x_a=\hat x_u} = \nabla_{x_u}\mathcal{H}|_{x_a=\bar z_2}=0_{m\times 1}$. Thus, $\hat x_a\in\cale$. Now it must be shown that $\nabla_{x_u} \calh(\psi(\hat x_a,\hat x_u,\bar d_u),\hat x_u)$ is equal to zero. By definition, this expression is equivalent to $\nabla_{x_u} \calh(\bar z_1,\bar z_2) = \nabla_{x_u} \calh(\bar x_a,\bar x_u)$, which is equal to zero by Assumption \ref{properAssump2}.
\end{IEEEproof}

\begin{IEEEbiography}[{\includegraphics[width=1in,height=1.25in,clip,keepaspectratio]{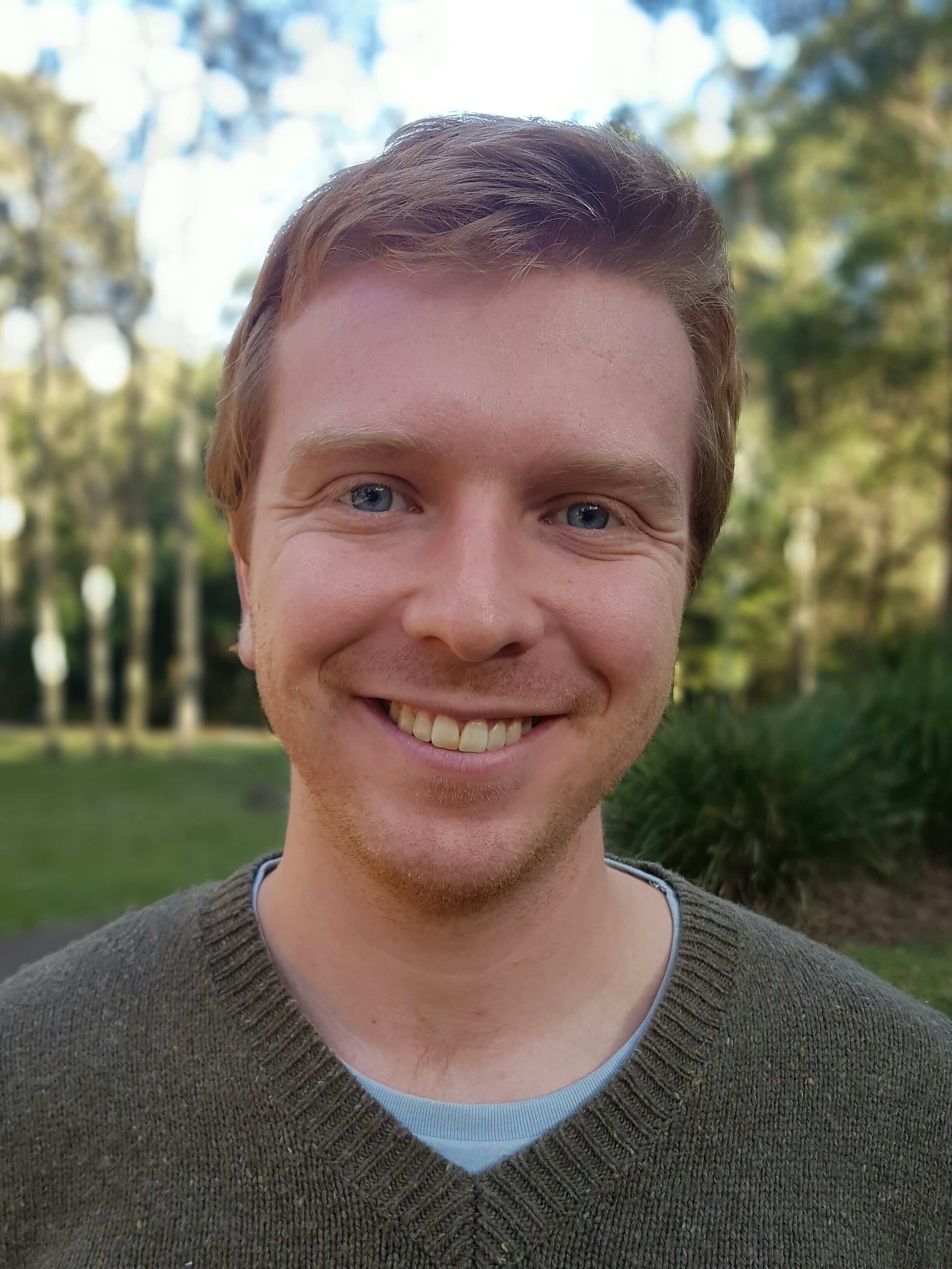}}]{Joel Ferguson}
was born in Australia. He received his degree in mechatronic engineering from the University of Newcastle in 2013, being awarded the Dean's medal and University medal. Since 2014, Joel has been pursuing his Ph.D. at the University of Newcastle under the supervision of Prof. R. Middleton and Dr. A. Donaire. His research interests include nonlinear control, port-Hamiltonian systems, nonholonomic systems and robotics.
\end{IEEEbiography}

\begin{IEEEbiography}[{\includegraphics[width=1in,height=1.25in,clip,keepaspectratio]{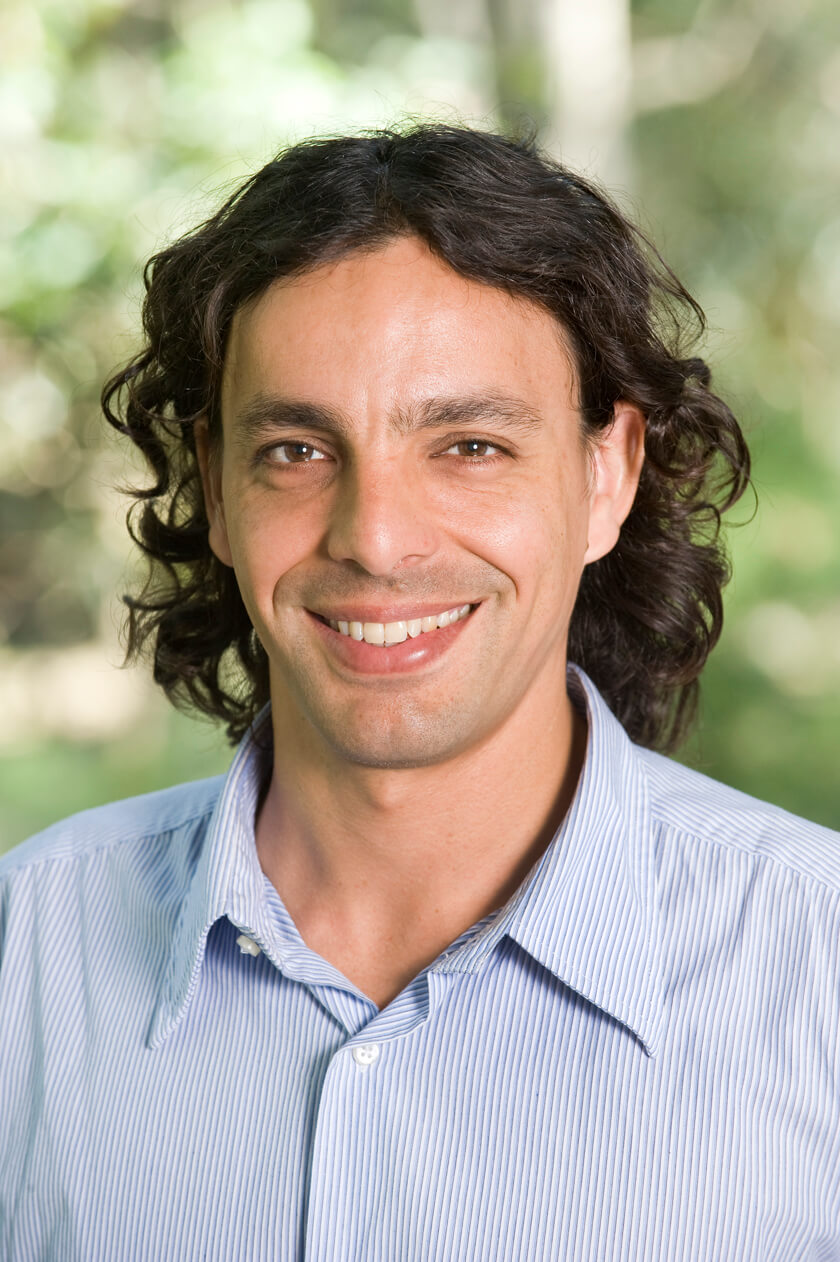}}]{Alejandro Donaire}
received his degree in Electronic Engineering in 2003 and his PhD in 2009, both from the National University of Rosario, Argentina. In 2009, he took a research position at the Centre for Complex Dynamic Systems and Control, The University of Newcastle, Australia. In 2011, he was awarded the Postdoctoral Research Fellowship at the University of Newcastle. In March 2015, he joined the robotic team at PRISMA Lab, University of Naples Federico II, Italy. Since 2017, he is conducting his research activities in within Queensland University of Technology (QUT), Australia. His research interests include nonlinear system analysis and passivity theory for control design of robotics, mechatronics, marine and aerospace systems.
\end{IEEEbiography}

\begin{IEEEbiography}[{\includegraphics[width=1in,height=1.25in,clip,keepaspectratio]{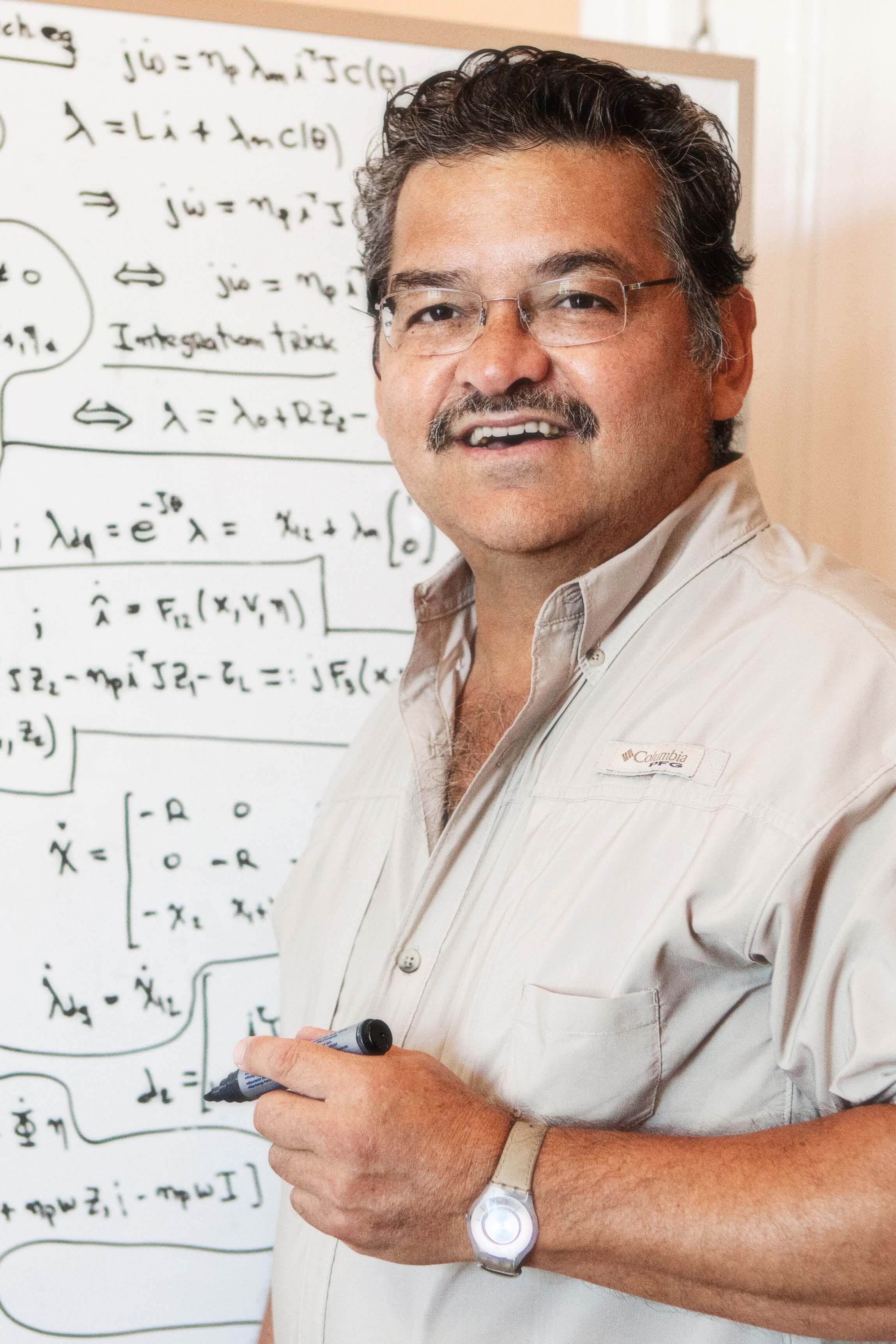}}]{Romeo Ortega}
was born in Mexico. He obtained his BSc in
Electrical and Mechanical Engineering from the National University of
Mexico, Master of Engineering from Polytechnical Institute of
Leningrad, USSR, and the Docteur D`Etat from the Politechnical
Institute of Grenoble, France in 1974, 1978 and 1984 respectively.

He then joined the National University of Mexico, where he worked until
1989. He was a Visiting Professor at the University of Illinois in
1987-88 and at McGill University in 1991-1992, and a Fellow of the
Japan Society for Promotion of Science in 1990-1991.  
He has been a member of the French National Research Council (CNRS) since June 1992.  Currently he is a Directeur de Recherche in the Laboratoire de Signaux et Systemes (CentraleSupelec) in Gif-sur-Yvette, France.  His research interests are in the fields of nonlinear and adaptive control, with special emphasis on applications.

Dr Ortega has published  three books and more than 290 scientific papers in international journals, with an h-index of 71. He has supervised more than 30 PhD thesis. He is a Fellow Member of the IEEE since 1999 and an IFAC Fellow since 2016. He has served as chairman in several IFAC and IEEE committees and participated in various editorial boards of international journals. 
\end{IEEEbiography}

\begin{IEEEbiography}[{\includegraphics[width=1in,height=1.25in,clip,keepaspectratio]{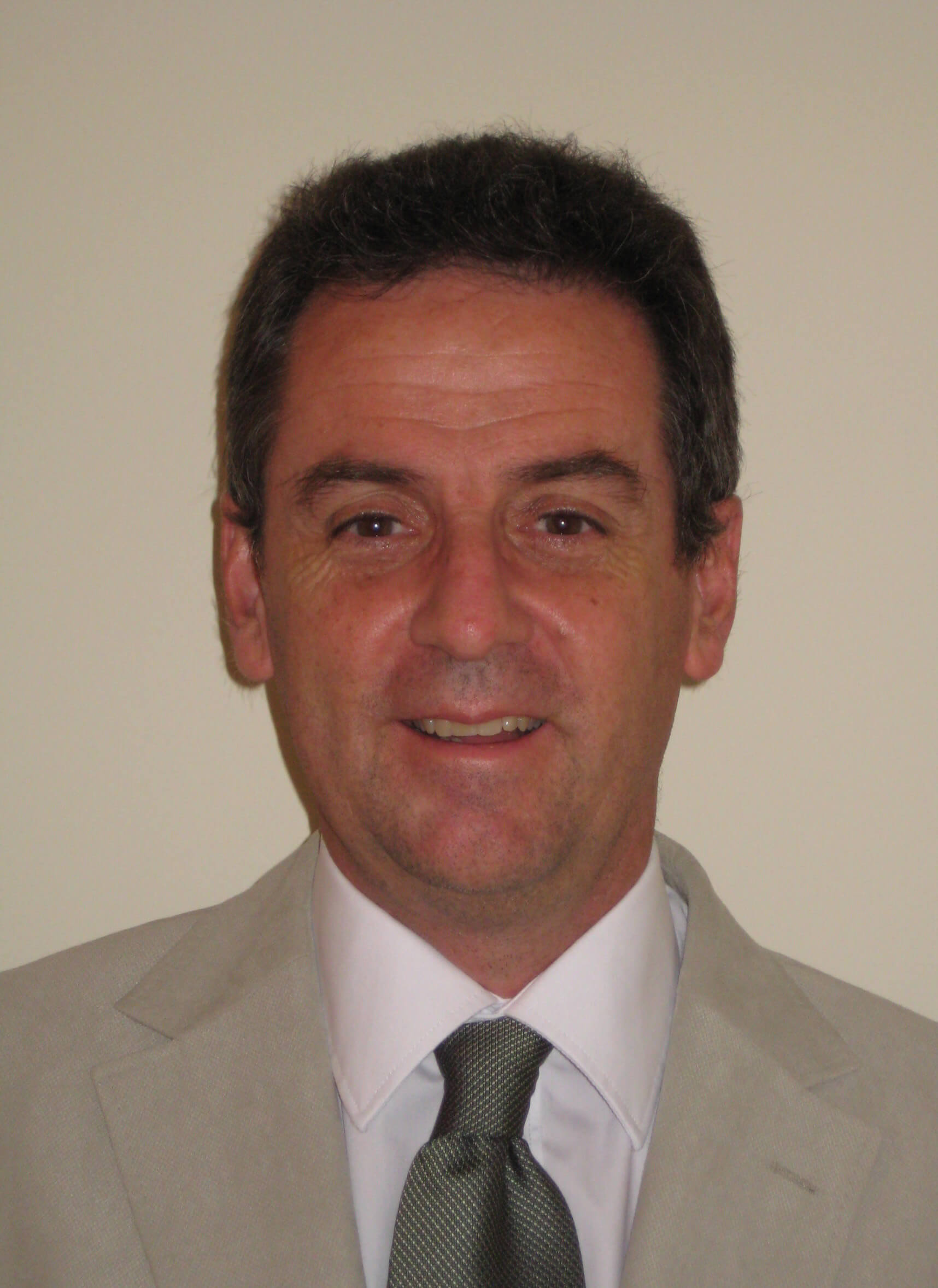}}]{Richard H. Middleton}
was born on 10th December 1961 in Newcastle Australia. He received his B.Sc. (1983), B.Eng. (Hons-I)(1984) and Ph.D. (1987) from the University of Newcastle, Australia. He has had visiting appointments at the University of Illinois at Urbana-Champaign, the University of Michigan and the Hamilton Institute (National University of Ireland Maynooth). In 1991 he was awarded the Australian Telecommunications and Electronics Research Board Outstanding Young Investigator award. In 1994 he was awarded the Royal Society of New South Wales Edgeworth-David Medal; he was elected to the grade of Fellow of the IEEE starting 1999, received the M.A. Sargent Award from the Electrical College of Engineers Australia in 2004, and as a Fellow of the International Federation of Automatic Control in 2013. 

He has served as an associate editor, associate editor at large and senior editor of the IEEE Transactions on Automatic Control, the IEEE Transactions on Control System Technology, and Automatica, as Head of Department of Electrical Engineering and Computing at the University of Newcastle; as a panel member and sub panel chair for the Australian Research Council; as Vice President - Member Activities and also as Vice President – Conference Activities of the IEEE Control Systems Society; President (2011) of the IEEE Control Systems Society;  as Director of the ARC Centre for Complex Dynamic Systems and Control; a Distinguished Lecturer for the IEEE Control Systems Society, and as a research professor at the Hamilton Institute, The National University of Ireland, Maynooth.

He is currently Head of School of Electrical Engineering and Computing at the University of Newcastle. He is also Editor – System and Control Theory, of Automatica. His research interests include a broad range of Control Systems Theory and Applications, including feedback performance limitations, information limited control and systems biology. 
\end{IEEEbiography}

\end{document}